\documentclass[journal,twoside,web]{ieeecolor}
\usepackage[utf8]{inputenc}
\usepackage{tmi}
\makeatletter
\def\ps@titlepagestyle{%
  \def\@oddfoot{}\def\@evenfoot{}%
  \def\@oddhead{\vbox{%
        \vbox{\begin{tabular}{@{}p{\textwidth}@{}}\\[-32pt]{\vbox{\hsize\textwidth\scriptsize\textsf\leftmark \hfill \textsf\thepage\hbox{}\par\vspace*{-3pt} 
        \vbox{\color{subsectioncolor}\hrule height1pt width\textwidth depth0pt} }}\end{tabular}} }}%
  \def\@evenhead{\begin{tabular}{@{}p{\textwidth}@{}}\\[-38pt]{\vbox{\hsize\textwidth\scriptsize\textsf\thepage \hfill \textsf\leftmark\hbox{}\par\vspace*{-3pt} 
        \vbox{\color{subsectioncolor}\hrule height1pt width\textwidth depth0pt} }}\end{tabular}\hskip-5pc\null
        }%
}
\makeatother
\usepackage{amsmath,amssymb,amsfonts}
\usepackage{algorithmic}
\usepackage{graphicx}
\usepackage{subfiles}
\usepackage{multirow}
\usepackage{booktabs}
\usepackage{hyperref}
\usepackage{textcomp}
\usepackage{cleveref}
\usepackage[style=ieee, maxbibnames=6, minbibnames=1]{biblatex}
\DeclareFieldFormat[article]{volume}{#1}
\DeclareFieldFormat[article]{number}{#1}

\renewbibmacro*{volume+number+eid}{%
  \printfield{volume}%
  \iffieldundef{number}{}{\mkbibparens{\printfield{number}}}%
  \setunit{\addcolon\space}
  \printfield{eid}%
}

\DeclareFieldFormat{pages}{#1}
\renewbibmacro*{pages}{\printfield{pages}}

\def\BibTeX{{\rm B\kern-.05em{\sc i\kern-.025em b}\kern-.08em
    T\kern-.1667em\lower.7ex\hbox{E}\kern-.125emX}}
\crefname{section}{Sec.}{Secs.}
\Crefname{section}{Sec.}{Secs.}
\crefname{figure}{Fig.}{Figs.}
\Crefname{figure}{Fig.}{Figs.}
\crefname{appendix}{App.}{Apps.}
\Crefname{appendix}{App.}{Apps.}

\newif\ifdraft
\drafttrue

\definecolor{orange}{rgb}{1,0.5,0}
\definecolor{pink}{rgb}{0.98, 0.38, 0.5}
\definecolor{darkgreen}{rgb}{0.055, 0.490, 0.016} 
\definecolor{amber}{rgb}{1.0, 0.75, 0.0}

\ifdraft
 \usepackage[normalem]{ulem}
 \newcommand{\FL}[1]{{\color{red}{\bf FL: #1}}}
 
 \newcommand{\EOE}[1]{{\color{orange}{\bf EOE: #1}}}
 
 \newcommand{\STM}[1]{{\color{blue}{\bf STM: #1}}}
 
  \newcommand{\MK}[1]{{\color{green}{\bf STM: #1}}}

\else
 \newcommand{\sout}[1]{}
 \newcommand{\FL}[1]{{\color{red}{}}}
 
 \newcommand{\EOE}[1]{{\color{orange}{}}}
 
 \newcommand{\STM}[1]{{\color{blue}{}}}
 
  \newcommand{\MK}[1]{{\color{green}{}}}
 
\fi

\begin{document}

\title{From Slices to Structures: Unsupervised 3D Reconstruction of Female Pelvic Anatomy from Freehand Transvaginal Ultrasound}

\author{Max Kr\"ahenmann, Sergio Tascon-Morales, Fabian Laumer, Julia E. Vogt, and Ece Ozkan 
\thanks{J.E. Vogt and E. Ozkan share last-authorship. }
\thanks{This work was supported by the InnoSuisse grant 119.406 IP-LS.}
\thanks{M. Kr\"ahenmann and E. Ozkan are with ETH Zurich, Universitaetstrasse 6, 8092, Zurich, Switzerland and University of Basel, Hegenheimermattweg 167C, 4123, Basel, Switzerland (e-mail: \{max.kraehenmann, ece.oezkanelsen\}@unibas.ch)}
\thanks{S. Tascon-Morales is with ETH Zurich, Universitaetstrasse 6, 8092, Zurich, Switzerland and Scanvio Medical AG, Zurich, Switzerland (e-mail: sergio.tasconmorales@inf.ethz.ch)}
\thanks{F. Laumer is with Scanvio Medical AG, Zurich, Switzerland (e-mail: fabian.laumer@scanvio.com). }
\thanks{J. E. Vogt is with ETH Zurich, Universitaetstrasse 6, 8092, Zurich, Switzerland (e-mail: julia.vogt@inf.ethz.ch). }}

\maketitle

\begin{abstract}
Volumetric ultrasound has the potential to significantly improve diagnostic accuracy and clinical decision-making, yet its widespread adoption remains limited by dependence on specialized hardware and restrictive acquisition protocols. 
In this work, we present a novel unsupervised framework for reconstructing 3D anatomical structures from freehand 2D transvaginal ultrasound sweeps, without requiring external tracking or learned pose estimators.
Our method, TVGS, adapts the principles of Gaussian Splatting to the domain of ultrasound, introducing a slice-aware, differentiable rasterizer tailored to the unique physics and geometry of ultrasound imaging. 
We model anatomy as a collection of anisotropic 3D Gaussians and optimize their parameters directly from image-level supervision. To ensure robustness against irregular probe motion, we introduce a joint optimization scheme that refines slice poses alongside anatomical structure. 
The result is a compact, flexible, and memory-efficient volumetric representation that captures anatomical detail with high spatial fidelity. 
This work demonstrates that accurate 3D reconstruction from 2D ultrasound images can be achieved through purely computational means, offering a scalable alternative to conventional 3D systems and enabling new opportunities for AI-assisted analysis and diagnosis.
\end{abstract}

\section{Introduction}
Three-dimensional (3D) ultrasound (US) imaging plays an essential role in obstetrics and gynecology, providing clinicians with a spatial context that is crucial for diagnosis, surgical planning, and treatment monitoring \cite{Alcazar2005,Jensen2007}. 
Transvaginal ultrasound (TVS) is widely used for high-resolution imaging of pelvic organs due to its affordability, safety, and minimally invasive nature, but extending 2D TVS to 3D imaging remains technically challenging.
Existing 3D US systems typically rely on external tracking mechanisms or mechanically swept probes. 
However, while 3D ultrasound imaging is commonly performed both transabdominally and transvaginally, especially in obstetric applications such as fetal imaging, most commercial 3D systems are designed for abdominal or general freehand use, not specifically for transvaginal gynecological contexts \cite{Apirakviriya2016,Yeung2024}.
In addition to being physically cumbersome and difficult to integrate into routine clinical workflows, these systems often suffer from lower resolution compared to conventional 2D imaging. 
The reliance on mechanical scanning or external tracking can introduce motion artifacts, particularly in dynamic or less controlled scanning environments. 
Furthermore, the need for specialized hardware increases system complexity and cost, which hinder widespread adoption.

Recent advances have explored machine learning-based methods for reconstructing 3D volumes from freehand 2D ultrasound sweeps. 
These approaches include pose estimation using speckle correlation, deep neural networks, and self-supervised trajectory inference \cite{Peng2022,Prevost2018,Eid2024}. 
While promising in specific domains, such methods are highly dependent on the training data, sensitive to imaging artifacts, and often lack generalizability across different acquisitions or anatomical regions. 
Current approaches still rely on generating pose information prior to reconstruction, which introduces an additional layer of complexity \cite{Eid2025}.

In this work, we propose \textbf{TVGS}: an unsupervised method for 3D US reconstruction that is free from the need for any external tracking or prior learned pose estimation. 
Our method is specifically designed for the gynecological domain and operates on standard 2D TVS sweeps acquired freehand. 
It leverages recent advances in volumetric rendering, adapting Gaussian Splatting \cite{Kerbl2023} to the constraints and physics of US imaging. 
Our approach uses sensorless geometric modeling of probe motion to establish spatial coherence between slices, and reconstructs anatomical structures using anisotropic Gaussian primitives optimized directly from US data.


Our framework represents anatomy as a collection of anisotropic Gaussians, optimized via a custom CUDA-optimized differentiable rasterizer. This continuous representation mitigates the aliasing and staircase artifacts inherent to discrete voxel-based methods, particularly in off-axis slices, while achieving convergence speeds significantly faster than implicit neural representations. Importantly, our method requires no calibration data and does not rely on training with large datasets.


We validate our method on real-world clinical TVS data, reconstructing 3D volumes from two orthogonal freehand sweeps. 
Despite being unsupervised and model-free, our approach produces anatomically accurate volumes that clearly capture key structures such as the uterus and endometrium, with strong visual and spatial fidelity. 
Our method also demonstrates robustness under realistic acquisition conditions and shows potential for real-time performance.
In summary, our contributions are:
\begin{itemize}
    \item An unsupervised 3D reconstruction method from 2D TVS sweeps, requiring no external tracking or learned pose estimators.
    \item A custom volumetric framework using Gaussian Splatting, adapted for US-specific imaging physics.
    \item A CUDA-optimized differentiable rasterizer tailored for  slice-based rendering under highly anisotropic sampling (dense in-plane, sparse through-plane).
    \item A joint optimization strategy that refines slice poses alongside scene parameters, increasing robustness to irregular probe motion typical in freehand acquisition.
\end{itemize}  
\section{Related Work}
2D to 3D US reconstruction has been approached using a variety of strategies. 
These approaches differ significantly in their reliance on external tracking hardware, training data, and assumptions about probe motion, each carrying implications for their suitability in clinical workflows, particularly in TVS.


Early systems using electromagnetic sensors \cite{Wen2013} provided accurate positioning but are ill-suited for TVS due to bulk and patient discomfort \cite{Peng2022}. Consequently, sensorless methods emerged, leveraging signal processing techniques like speckle decorrelation \cite{Chen1997, Gee2006, Li2002} and spatial correlation \cite{Gilliam2006} to infer probe movement. While foundational, these approaches often struggle with noise and complex multi-axis motion. Later statistical and graph-based optimizations \cite{Tetrel2016} improved robustness but remained too fragile for widespread clinical adoption.

With the rise of deep learning, supervised learning methods for pose estimation gained popularity. 
These approaches train convolutional or recurrent neural networks to predict either inter-frame transformations or directly reconstruct 3D volumes. 
Early models demonstrated that CNNs could infer relative probe motion from US frames, enabling trackerless 3D reconstruction \cite{Prevost2018}. 
Later methods like $\text{DC}^2$-Net \cite{Guo2023} introduced contrastive and temporal losses to improve trajectory consistency across sequences.
Another noteworthy model is PLPPI, which incorporates domain-specific physics constraints into the learning pipeline, enabling improved handling of out-of-plane motion \cite{Dou2024}. 
However, a shared limitation of these models is their dependency on ground truth data from tracked acquisitions for training and evaluation. 
In many clinical workflows—particularly for TVS—such tracking data is unavailable or infeasible to collect.
Some recent works like RecON \cite{Luo2023} explore online learning, updating models in real-time as new frames are acquired. 
While promising for reducing drift and latency, these methods still presuppose access to prior 3D volumes or strong priors, making them unsuitable for fully unsupervised clinical settings.

In parallel, implicit representations like Neural Radiance Fields (NeRF) \cite{Mildenhall2020} and subsequent developments (e.g., ImplicitVol \cite{Yeung2024}) model 3D structures as continuous functions learned by neural networks. 
These models offer resolution independence and are memory-efficient but typically require extensive training and known camera poses—conditions that are generally not met in US unless a tracking device is present.
Hybrid methods such as RapidVol \cite{Eid2024} attempt to blend implicit and explicit approaches, using tri-planar maps refined by lightweight networks to accelerate training and inference. 
However, even these models typically assume access to 3D volumes for training or initialization.

Beyond motion estimation, the choice of volumetric representation plays a central role in reconstruction quality and efficiency. 
Traditional explicit methods such as voxel grids offer high interpretability and easy integration into existing medical imaging pipelines. 
However, they suffer from cubic memory growth with resolution and are thus computationally intensive for fine anatomical detail.
More recently, techniques like Gaussian Splatting have emerged as a powerful alternative. 
Originally developed for real-time rendering in computer vision \cite{Kerbl2023}, Gaussian Splatting represents scenes using parameterized Gaussian primitives whose locations, shapes, and radiance are optimized to match rendered views. 
UltraGauss \cite{Eid2025} adapted this to fetal US imaging by incorporating anisotropic Gaussians and integrating pose predictions from QAERTS \cite{Ramesh2024}. 
This approach improves rendering quality and speed but still depends on learned pose models, which hinders generalization and deployment in tracking-free environments.

In contrast to these existing approaches, our method introduces the first application of Gaussian Splatting for 3D US that is entirely unsupervised and sensorless. 
We eliminate the need for pose estimation models, supervision, or tracking hardware by leveraging geometric assumptions on sweep motion and optimizing anisotropic Gaussian primitives directly from 2D TVS slices. 
This enables anatomically accurate, scalable 3D reconstruction using only standard 2D US equipment, significantly lowering the barrier to clinical deployment and advancing the capabilities of low-cost US imaging.
\section{Datasets}
\label{sec:dataset}

To develop and evaluate our unsupervised 3D reconstruction framework, we utilize both real and synthetic TVS data. 
Synthetic data offers controlled conditions with ground-truth pose labels for benchmarking, while real clinical sweeps provide representative anatomical complexity and scanning variability. 

\subsection{Synthetic Data}
\begin{figure}[h]
    \centering
    \includegraphics[width=0.8\linewidth]{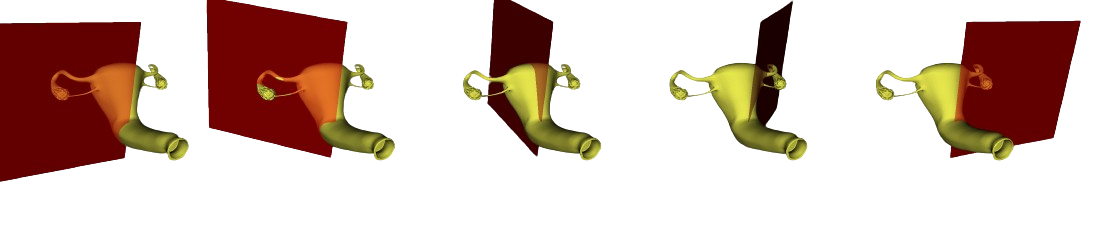}\\
    \includegraphics[width=0.8\linewidth]{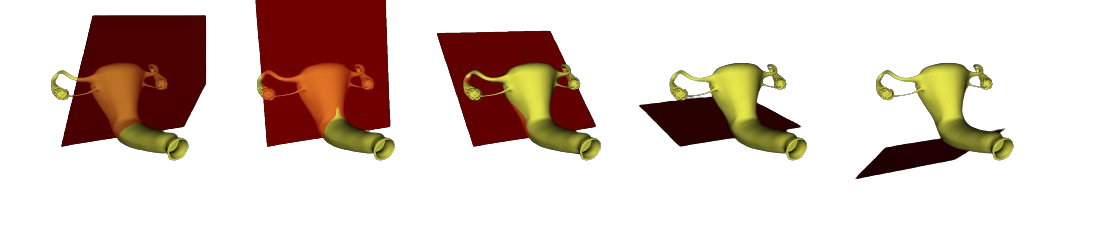}
    \caption{\textit{Top:} Simulated sagittal sweep. \textit{Bottom:} Simulated transversal sweep. The image only shows some of the generated probe rotations.}
    \label{fig:uterus}
\end{figure}

To support development and controlled evaluation, we construct a synthetic dataset. 
Starting from a 3D uterus mesh in STL format, we voxelized the model into a segmentation volume with three labels: interior ($0.5$), border ($1$), and exterior ($0$). 
Using 3D Slicer \cite{Fedorov2012}, slices are generated by simulating probe rotations through angular sweeps, ranging from $-60^\circ$ to $+60^\circ$ in sagittal and transversal directions, as shown in \Cref{fig:uterus}.
To reflect realistic acquisition scenarios, the slices are sparsely sampled along each sweep, with a limited number of angular steps per direction, resulting in a coarse and uneven spatial coverage of the volume.

\begin{figure}[h]
    \centering
    \includegraphics[width=.2\linewidth]{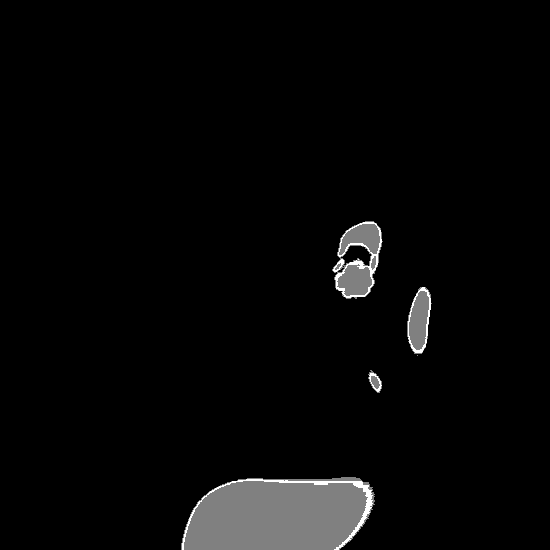}
    \includegraphics[width=.2\linewidth]{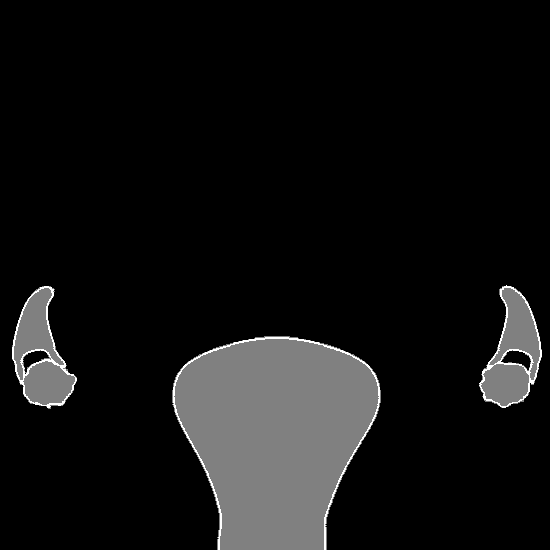}
    \caption{
        Synthetic ultrasound slices generated from a simulated uterus model using a renderer. 
        \textit{Left}: Sagittal view showing longitudinal anatomy. 
        \textit{Right}: Transversal view perpendicular to the sagittal plane. These slices can be sampled at arbitrary resolution.
    }
    \label{fig:synthetic_frames}
\end{figure}

Each slice is associated with a known 6D pose vector:
\begin{equation}
\mathbf{y} = [r_x, r_y, r_z, t_x, t_y, t_z],
\end{equation}
where $\mathbf{r}$ and $\mathbf{t}$ denote Euler angles and translations with respect to a canonical coordinate system.
This ground truth facilitates quantitative evaluation of slice pose and 3D reconstruction accuracy.
Examples of generated frames are shown in Fig. \ref{fig:synthetic_frames} for sagittal (left) and transversal (right).

\subsection{Real Data}
\label{sec:real}
\begin{figure}
    \centering
    \includegraphics[trim=0 0 0 8.8cm, clip, width=0.35\linewidth]{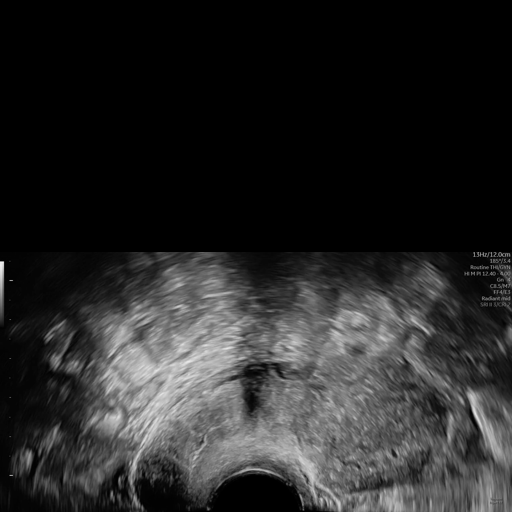}
    \includegraphics[trim=0 0 0 8.8cm, clip, width=0.35\linewidth]{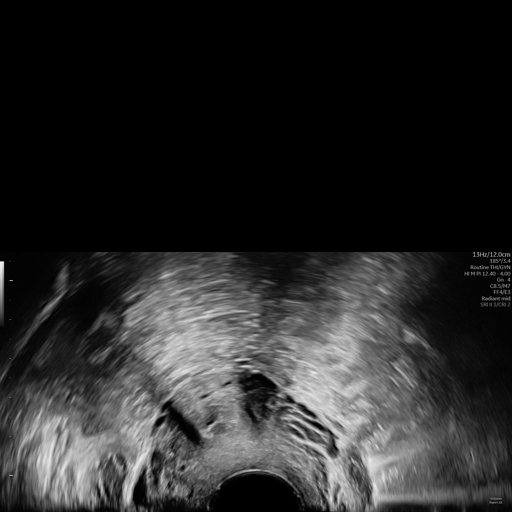}
    \caption{Two frames sampled from the real data sweeps. On the left, a frame taken from the sagittal view. On the right, a frame taken from the transversal view.}
    \label{fig:real}
\end{figure}

20 TVS sweeps were acquired from 10 patients (2 sweeps per patient) using a GEHC Voluson Expert 22 ultrasound machine. 
\Cref{tab:real_data_params} summarizes the main parameters of the data. 
Each sweep represents either a sagittal or transversal view, with probe motion approximated as a smooth angular trajectory. 
An example image for both modalities can be seen in \Cref{fig:real}. 
Each patient provided written informed consent prior to participation, and all procedures adhered to strict ethical standards to ensure confidentiality and privacy.
Since real data lacks tracked probe poses, we assume consistent frame intervals and smooth probe motion. We initialize the pose of each frame using equal angular spacing. 
This assumption mirrors the angular sampling strategy used in generating synthetic data. 
Other than resizing and padding images when needed, no additional transformations or data augmentation techniques are applied, ensuring that the intrinsic properties of the images are preserved.
Our method assumes (i) temporally ordered slices with approximately uniform angular spacing, and (ii) minimal tissue deformation during the sweep.
These assumptions are critical for effective sensorless reconstruction, and we analyze their impact in Section~\ref{sec:unsupervised_gaussian_splatting}.

\begin{table}[b]
    \centering
    \caption{Main parameters of real data. Values in parentheses represent the standard deviation.}
    \label{tab:real_data_params}
    \begin{tabular}{cl}
        \toprule
        \textbf{Parameter} & \textbf{Value(s)} \\
        \midrule
        Frames per second & 10-20 \\
        Duration & $\sim$10 seconds \\
        Frames per sweep & 100-300 \\
        Plane & Sagittal / Transversal  \\
        Original Resolution & 1528 x 784\\
        \midrule
        Age & 41.9 (6.0) \\
        Height [m] & 1.67 (0.07)\\
        Weight [kg] & 68.49 (13.26)\\
        Ethnicity & White/European\\
        \bottomrule
    \end{tabular}
\end{table}

\subsection{Test Set}

For quantitative evaluation, we establish a distinct test set partition. Although the raw sweep lengths vary, we standardize our evaluation by subsampling 85 equidistant slices from each sweep. From these, we select 10 slices uniformly at random to serve as the test set and keep the remaining 75 for training (we analyze the impact of different slice counts in Section \ref{subsec:slice_density}). Crucially, because the ultrasound slices represent thin, planar cross-sections with sparse angular spacing, these held-out test slices do not spatially overlap with any of the remaining slices used for training. Consequently, this evaluation setup is particularly challenging; it effectively constitutes a task of novel view synthesis (NVS), requiring the model to accurately infer anatomical structures in the gaps between training views without direct supervision. As we do not have accurate poses for NVS, we perform a short optimization of them before evaluation to factor out pose error, consistent with standard practices for pose estimation in neural rendering frameworks \cite{Lin2021}.
\section{Unsupervised Gaussian Splatting for Ultrasound}
\label{sec:unsupervised_gaussian_splatting}
\subsection{Overview: From Gaussian Splatting to Ultrasound Imaging}

Gaussian Splatting is a recent real-time rendering method that represents a 3D scene using volumetric Gaussian primitives~\cite{Kerbl2023}. 
Each Gaussian contributes to pixels in projected views based on its position, shape (covariance), opacity, and radiance. 
The method is differentiable and well-suited for learning-based optimization, particularly in settings where redundant multi-view image data is available (e.g., RGB images from moving cameras).

However, US imaging presents fundamental differences. 
Each image (or ``slice") corresponds to a narrow, planar cross-section of tissue, implying limited overlap between any two slices. In TVS, this limitation is even more pronounced as there is no overlap between slices in the sweep.
As a result, there is no redundancy between views to help maintain scene consistency.
Additionally, the physics of US introduces directionality and discontinuities that standard perspective projection does not capture. This stands in contrast to Gaussian Splatting for optical images, where we work with projections of the scene onto a camera sensor, inherently providing depth information. Since US does not have this projection component and instead captures direct cross-sections, approaches relying on projection geometry cannot be directly applied.
Misaligned Gaussian primitives—especially in depth—can significantly affect the rendering, since each slice is extremely thin.
To address these challenges, we adapt Gaussian Splatting, building a slice-aware differentiable rasterizer specifically designed for US data. 
Our framework accounts for the sparse and largely non-overlapping slices while retaining the strengths of the Gaussian formulation—compact, continuous scene representation and differentiability.

\begin{figure}
    \centering
    \includegraphics[width=\linewidth]{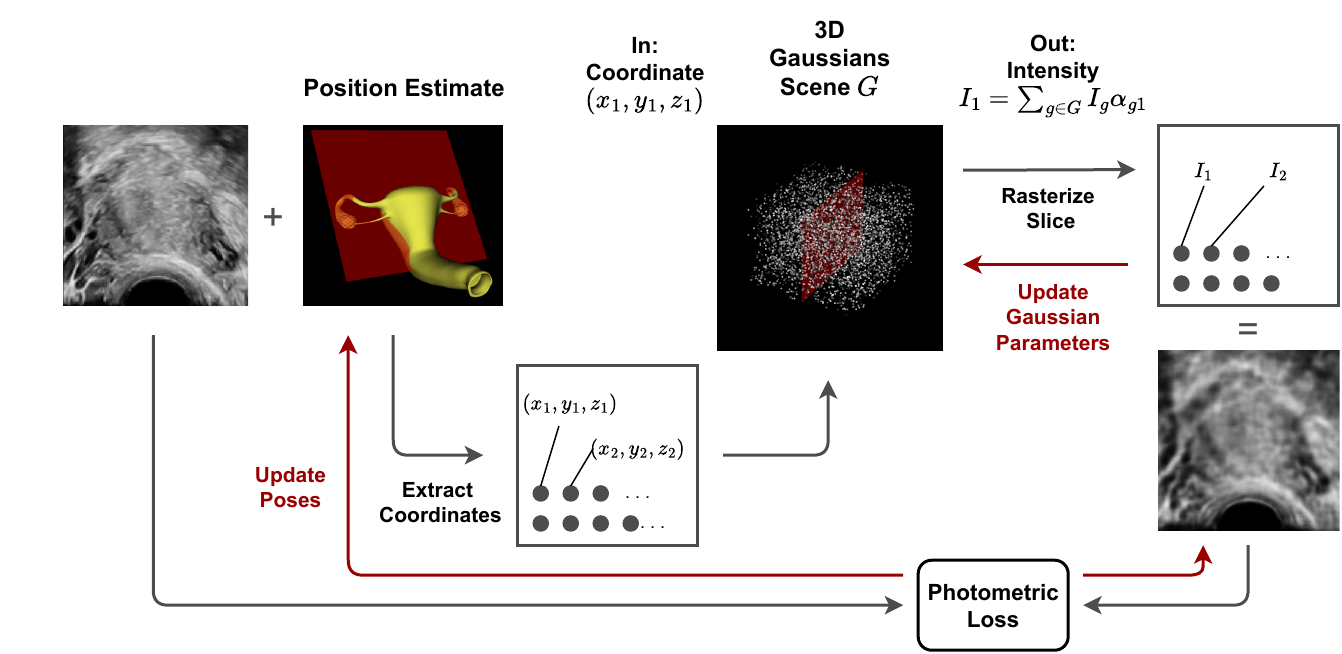}
    \caption{
        Overview of our differentiable Gaussian Splatting pipeline for ultrasound volume reconstruction. 
        A set of 3D Gaussians—each defined by a position, shape, intensity, and opacity—is rendered onto input ultrasound slice planes via a custom rasterizer. 
        A reconstruction loss drives gradient-based optimization of all Gaussian parameters. The framework includes strategies for initialization, pruning, and performance optimization.
    }
    \label{fig:gs}
\end{figure}

An overview of our method is shown in \Cref{fig:gs}. 
We represent anatomy as a set of volumetric 3D Gaussians, which are rendered onto slice planes using a custom differentiable rasterizer. 
This enables end-to-end optimization of all parameters using only image-level supervision from sparse slices.

\subsection{Slice-Based Differentiable Rasterization}
\label{sec:forward}

Our rasterizer computes the intensity of each pixel in a slice by aggregating contributions from a set of 3D Gaussian primitives. 
We begin by describing the formulation in terms of its constituent steps.
Each Gaussian \( g \) is defined by a set of parameters:
\begin{itemize}
    \item \( \mathbf{\mu}_g \in \mathbb{R}^3 \): the center of the Gaussian,
    \item \( \Sigma_g \in \mathbb{R}^{3 \times 3} \): the covariance matrix describing its shape and orientation. It is factored in the same way as in the original GS formulation \cite{Kerbl2023}: \( \Sigma_g = (S_g R_g)^\top (S_g R_g) \), where \( S_g \) is a diagonal scale matrix and \( R_g \) is a rotation matrix derived from a unit quaternion,
    \item \( o_g \in [0,1] \): the opacity, controlling transparency,
    \item \( I_g \in \mathbb{R} \): the scalar intensity value associated with the Gaussian.
\end{itemize}

Each pixel \( p \) in a US slice is associated with a known 3D coordinate \( \mathbf{c}_p \in \mathbb{R}^3 \), obtained via slice pose modeling as described in Section \ref{sec:real} and the location of the pixel in the slice. Notably, $\mathbf{c}_p$ is a differentiable function of the slice's rigid body transformation parameters. Consequently, gradients computed at the pixel level during the backward pass flow not only to the Gaussian parameters but also to the extrinsic matrix of the slice, enabling the correction of misalignment errors.
For each Gaussian \( g \), we compute the offset vector from the Gaussian center to the pixel coordinate:
\begin{equation}
    \mathbf{d}_{gp} = \mathbf{c}_p - \mathbf{\mu}_g, 
\end{equation}
This displacement indicates how far and in which direction pixel \( p \) lies from the Gaussian center.
We use the Mahalanobis distance to compute how ``close" the pixel is to the Gaussian in the metric induced by the covariance:
\begin{equation}
    e_{gp} = -\tfrac{1}{2} \mathbf{d}_{gp}^\top \Sigma_g^{-1} \mathbf{d}_{gp}.
\end{equation}
This term governs the spatial falloff of the Gaussian’s influence. 
A higher value of \( e_{gp} \) (closer to 0) means greater contribution.
The influence of Gaussian \( g \) at pixel \( p \) is then:
\begin{equation}
    \alpha_{gp} = o_g \exp(e_{gp}),
\end{equation}
where \( o_g \) scales the exponential according to the visibility (opacity) of the Gaussian.
The final intensity \( I_p \) at pixel \( p \) is computed as a weighted sum of intensities from all Gaussians:
\begin{equation}
    I_p = \sum_g \alpha_{gp} I_g.
\end{equation}

This formulation defines each pixel as the sum of soft and spatially-varying influences from 3D Gaussians. 
By using Gaussian functions instead of hard surface projections, the method can represent continuous, soft-tissue-like structures and remains differentiable for optimization. 
It is especially well-suited for sparse data with anisotropic spatial sampling such as US slices, where direct projection is unreliable and information across slices is limited.

To enable efficient end-to-end optimization of this representation, we implement the rasterizer as a high-performance CUDA kernel.
The custom nature of this slice-accumulation process precludes the use of standard automatic differentiation frameworks like PyTorch's \texttt{autograd}.
Instead, we analytically derive the gradients for all Gaussian parameters and implement them directly within the kernel using the chain rule.
This manual backward pass also avoids heavy automatic differentiation graphs, resulting in significant memory savings.

\subsection{Parameter Initialization}
\label{sec:initialization}
Effective initialization of Gaussian parameters is crucial to ensure stable and efficient optimization. 
This is particularly important because the contribution of each Gaussian to the rendered image, thus, the magnitude of its gradient updates—depends multiplicatively on both its intensity \( I_g \) and opacity \( o_g \). 
If these parameters are poorly initialized (e.g., near zero), the Gaussian becomes effectively invisible and receives negligible gradient updates, rendering it ``inactive" from the outset.
If initialized randomly, we could potentially run into such issues; the following values were found to work well empirically for our setting:
\begin{itemize}
    \item For moderate initial opacity, $o_g = \sigma(1.0) \approx 0.73$,
    \item For mid-range intensity, $I_g = 0.5$,
    \item For moderate spatial extent, $\mathbf{s}_g = \exp([0.5, 0.5, 0.5])^\top \approx [1.65, 1.65, 1.65]^\top$
    \item For identity rotation, $\mathbf{q}_g = [1, 0, 0, 0]^\top$.
\end{itemize}
This choice of initialization ensures that each Gaussian has a non-negligible influence on the rendered slice at the start of training, gradients with respect to spatial and shape parameters are not suppressed, and training begins from a smooth, isotropic configuration, avoiding early instability from sharp or ill-posed Gaussians.
Unlike opacity and intensity, the mean position \( \mathbf{\mu}_g \in \mathbb{R}^3 \) is highly sensitive to initialization. 
Gaussians must be spatially close to the slices they are meant to influence; otherwise, their contribution to the rendered image (and thus their gradient signal) is exponentially suppressed via the term \( \exp(e_{gp}) \). 
In practice, we found that poor initialization of \( \mathbf{\mu}_g \) leads to Gaussians becoming permanently inactive before they can migrate toward meaningful regions.
We consider two strategies for initializing \( \mathbf{\mu}_g \):

\textit{- On-slice initialization}: Given a set of slice poses—whether inferred via our sensorless motion model or obtained from external tracking—we distribute Gaussian centers uniformly across the image planes. This places primitives directly within the signal domain from the start, ensuring strong initial gradient updates and accelerating convergence.
    
\textit{- Uniform sampling in a bounding volume}: Alternatively, we sample Gaussian centers uniformly within a predefined 3D bounding box enclosing the volume of interest. While this strategy is agnostic to the initial slice configuration, it typically requires additional optimization iterations for primitives to migrate from empty space toward informative anatomical regions.

\subsection{Inactive Gaussians and Density Control}
\label{sec:density}
During training, certain Gaussians may become effectively ``inactive"—typically those initialized in regions of the volume that are consistently dark or far from any informative US data. 
These Gaussians tend to receive low gradient signals, leading to rapid decay of their opacity \( o_g \) and intensity \( I_g \) toward zero. 
Once these values are sufficiently small, the Gaussian's contribution to the image becomes negligible, which in turn suppresses all subsequent gradient updates to its position, shape, and rotation. 
In effect, these Gaussians are permanently frozen in an uninfluential state.

Such inactive Gaussians waste computational resources. 
To mitigate this, we incorporate a lightweight density control mechanism that periodically removes such Gaussians and replaces them with new ones better positioned to contribute meaningfully.

We define a simple but effective scalar activity metric:
\[
m_g = |I_g \cdot o_g|
\]
Gaussians for which \( m_g < \epsilon \) (where \( \epsilon \) is a low predefined threshold) are considered inactive. 
These are removed from the parameter set every \( N \) optimization steps. 
This process is efficient to implement and avoids the overhead of evaluating gradient norms or rendering error sensitivity for  Gaussians.

To maintain volumetric capacity, we re-seed an equivalent number of new Gaussians, sampling their means uniformly within the bounding box of currently active primitives ($\mathbf{\mu}_{\text{new}} \sim \mathcal{U}[\mathbf{\mu}_{\min},\, \mathbf{\mu}_{\max}]$). 
This simple heuristic leverages the existing distribution of active Gaussians to guide new candidates toward potentially informative regions, without requiring a full analysis of image-space reconstruction error.
Other parameters are reset to default values as described in \Cref{sec:initialization}.

As a complementary strategy, we apply a non-uniform learning rate schedule where the Gaussian means \( \mathbf{\mu}_g \) are updated more aggressively than other parameters (e.g., opacity, intensity, and rotation). 
The motivation is to allow Gaussians to quickly migrate toward regions where they influence the rendered image before their visibility can be ``shut off" by the optimizer. 
In more extreme variants, we temporarily freeze non-positional parameters entirely during early training epochs. 
This prevents premature deactivation and leads to more effective exploration of the volume.

Both the pruning–reinitialization mechanism and learning rate scheduling are essential for maintaining a healthy and responsive population of Gaussians during training. 

\subsection{Joint Pose and Scene Optimization}
\label{sec:pose_opt}
A core challenge in sensorless freehand ultrasound is the variability in probe motion. While we initialize our reconstruction assuming a smooth, linear angular sweep, human hand motion inevitably introduces irregularities. To address this, we decouple the slice poses from the fixed geometric prior.
We treat the pose parameters $\mathbf{y}_i = [\mathbf{r}_i, \mathbf{t}_i]$ for each slice $i$ as learnable parameters. During the backward pass, gradients are propagated through the rasterizer not only to the Gaussian primitives but also to the transformation matrices defining the slice positions. This allows the model to fine-tune the alignment of individual slices to maximize photometric consistency with the reconstructed volume. We apply a lower learning rate to pose parameters compared to scene parameters to prevent drift and maintain the global structure defined by the initialization.

\subsection{Practical Considerations for Implementation}
\label{sec:implementation_summary} 
To make training tractable at scale, we apply a series of CUDA-level optimizations across both the forward and backward passes of the differentiable rasterizer. 
In particular, we precompute invariant quantities such as inverse covariance matrices ($\Sigma_g^{-1}$), which can then be reused across all affected pixels. 
We also adopt shared memory tiling, where Gaussian parameters are loaded once per tile and reused by all threads in a block, thereby minimizing global memory traffic. 
Further efficiency is gained through symmetry-aware matrix handling, which exploits the symmetric structure of covariance matrices to reduce both memory usage and compute overhead, and through matrix-free computation strategies that replace expensive general-purpose matrix operations with lightweight element-wise arithmetic. 
Together, these optimizations improve training throughput by an order of magnitude compared to a naive implementation and are critical for scaling to hundreds of thousands of Gaussians in high-resolution 3D volumes.

\section{Experiments and Results}

\subsection{Experimental Setup}
We followed standard practices for image resolution and batch size, while empirically tuning remaining settings to ensure high fidelity. Unless otherwise specified, we train for 150 epochs with a batch size of 32. We use a resolution of $256\times256$ for synthetic data and $256\times128$ for real data, the latter selected to match the dataset's native 2:1 form factor. The scenes are represented with approximately 50k Gaussians, and optimized using a hybrid loss ($\lambda_\text{L1}=0.8, \lambda_\text{SSIM}=0.2$) with learning rates 0.2 (means), 0.03 (opacity), 0.01 (scale), and 0.008 (intensity). The selection of specific values is explained in more detail in \Cref{subsec:optimizations}. We divide the experiments into optimizations (\Cref{subsec:optimizations}) and reconstruction quality (\Cref{subsec:reconstruction_quality}).
To comprehensively evaluate model performance, we report standard image similarity metrics: Structural Similarity Index Measure (SSIM), Peak Signal-to-Noise Ratio (PSNR), and Learned Perceptual Image Patch Similarity (LPIPS), using their conventional formulations.

\subsection{Optimizations}
\label{subsec:optimizations}
In this section, we evaluate the impact of various method configurations on reconstruction performance. First, we explored different training objectives, testing L1, L2, SSIM, PSNR, NCC, LPIPS and hybrid combinations. We found that a hybrid configuration ($\lambda_\text{L1}=0.8, \lambda_\text{SSIM}=0.2$) yielded the best results, offering superior robustness across metrics compared to the other tested functions. Consequently, we select this loss for all subsequent experiments. 

\subsubsection{Results of Optimization Techniques}
\begin{figure}
    \centering
    \includegraphics[width=.35\linewidth]{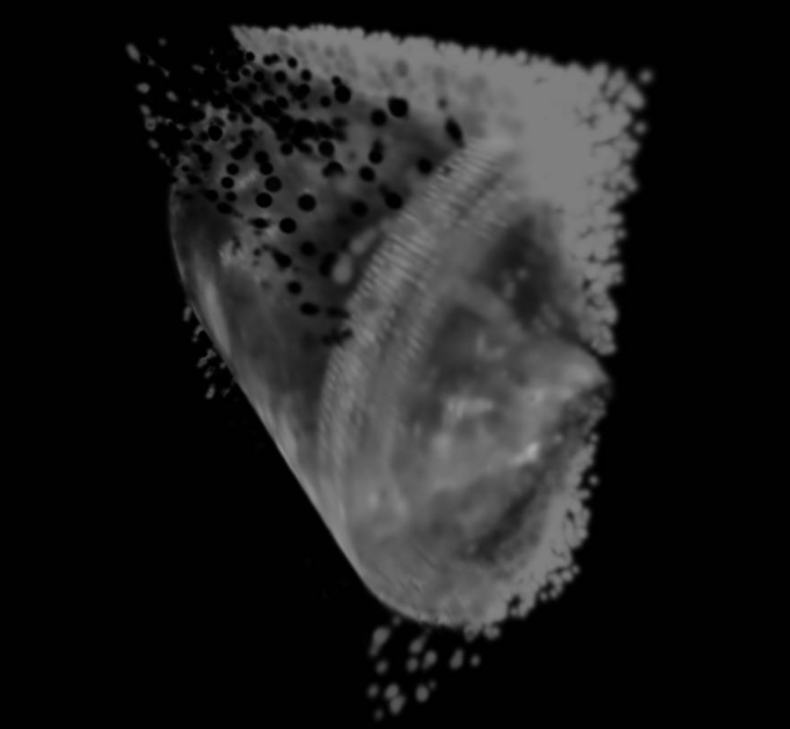}
    \caption{Example of a reconstructed 3D view scene after uniform random initialization of means.}
    \label{fig:example_initialization_on_slice}
\end{figure}

\paragraph{Gaussian Mean initialization} 
To evaluate the suitability of each of the initialization approaches described in \Cref{sec:initialization}, we run experiments and report the SSIM, PSNR and LPIPS for each case, as shown in \Cref{tab:results_mean_initialization}. \Cref{fig:example_initialization_on_slice} shows an example of the resulting scene obtained through uniform random initialization. 

\begin{table}[ht]
    \centering
    \caption{SSIM, PSNR and LPIPS metrics on runs with different initialization for sagittal and transversal view optimization.}
    \label{tab:results_mean_initialization}
    \begin{tabular}{l l c c c}
        \toprule
        \textbf{Group} & \textbf{Init. method} & SSIM ($\uparrow$) & PSNR ($\uparrow$) & LPIPS($\downarrow$)\\
        \midrule
        \multirow{2}{*}{Sagittal} & Random & 0.960 & 26.49 & 0.083\\
                                  & On-Slice & \textbf{0.975} & \textbf{28.54} & \textbf{0.057}\\ 
        \midrule
        \multirow{2}{*}{Transversal} & Random & 0.974 & 28.25 & 0.062\\
                                  & On-Slice & \textbf{0.983} &  \textbf{30.38} & \textbf{0.047}\\
        \bottomrule
    \end{tabular}
\end{table}

As shown in \Cref{tab:results_mean_initialization}, on-slice initialization yields consistently superior metrics and is selected as the default method. We attribute this performance gap to the sparse nature of US slice data. Random initialization frequently places primitives in empty space far from the image planes; these primitives contribute negligible density to the rendering, resulting in vanishing gradients that prevent them from migrating toward anatomical structures. In contrast, on-slice initialization guarantees that primitives spatially overlap with valid signal from the first iteration. This ensures immediate gradient flow and prevents primitives from becoming effectively ``inactive" at the start of training—a challenge we analyze further in the following section on density control.

\paragraph{Number of Gaussians}
The number of Gaussian primitives directly influences both reconstruction quality and computational cost. Increasing this number provides the model with more flexibility to capture fine anatomical details, potentially enhancing reconstruction fidelity. However, this improvement comes with a higher memory footprint and significantly longer training times.

\begin{figure}
\centering
\includegraphics[width=\linewidth]{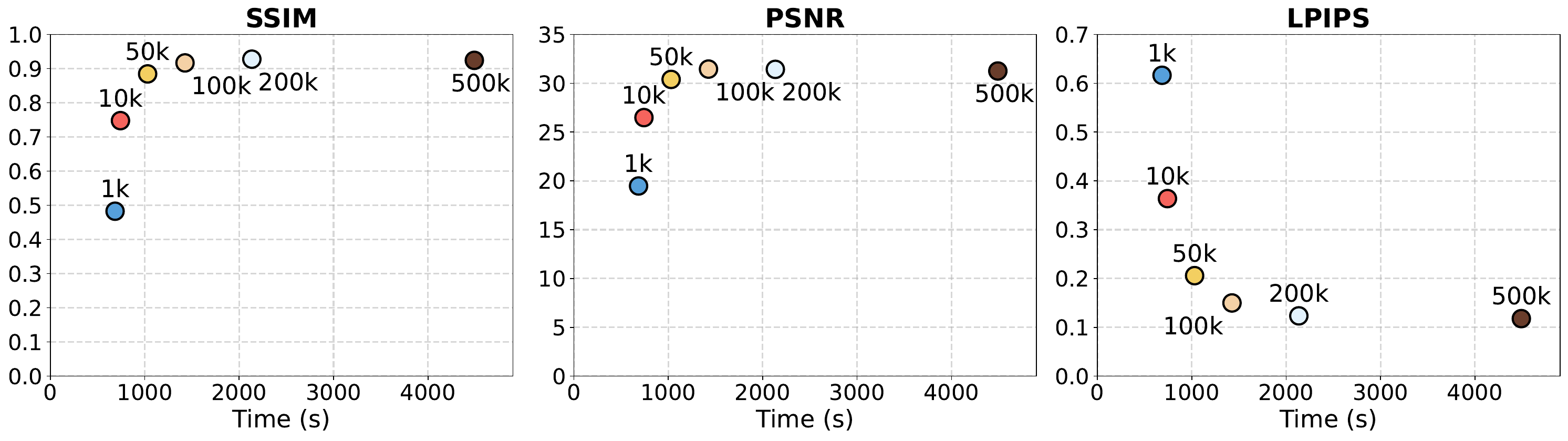}
\caption{Mean SSIM, PSNR and LPIPS vs. training time for different numbers of Gaussians representing the scene.}
\label{fig:gs_res}
\end{figure}

To explore this trade-off, we systematically varied the number of Gaussians on a fixed set of real data sweeps. We monitored reconstruction quality metric alongside the total training time, as shown in \Cref{fig:gs_res}. As expected, a higher number of Gaussians generally leads to improved reconstruction quality, reflected in lower L1 loss and higher SSIM scores. At the same time, training time increases substantially due to the added computational burden per iteration.

For the tested 256x128 resolution images, the results suggest an optimal range between approximately 60,000 Gaussians (about 600 per slice in the test setup) and 100,000 Gaussians (about 1000 per slice), which appears to strike a good balance between reconstruction quality and computational efficiency. Beyond roughly 100,000 Gaussians, the gains in L1 and SSIM begin to plateau, indicating diminishing returns, while the execution time continues to rise significantly. To strike a good balance between performance and execution time during experiments, we use approximately 50k Gaussians.

\paragraph{Inactive Gaussians}

As described before, inactive Gaussians can limit the effect of a large number of primitives on the final reconstruction. 
We apply reinitialization of Gaussians with spatial awareness, as described in \Cref{sec:density}, using different thresholds for pruning ($m_g = |I_go_g|$ values of 0.2, 0.1, 0.05, and 0.01) and test several schedules for when to apply the removal and addition steps. The best results were obtained using a threshold of 0.05 and applying density control every 25 epochs, starting at epoch 50 and continuing until epoch 125.

We observed that this mechanism introduces transient spikes in the loss function immediately following each pruning and reseeding step, though the optimization stabilizes rapidly. Quantitatively, the strategy yields minor improvements, primarily when using random initialization where the risk of inactive primitives is higher. In this setting, applying density control improved SSIM from 0.974 to 0.978, PSNR from 28.25 dB to 28.82 dB, and LPIPS from 0.062 to 0.053. Thus, while not strictly necessary for convergence—especially given the robustness of our default on-slice initialization—it serves as a useful refinement to maximize performance.

\subsubsection{Robustness to Motion Jitter}
Real-world sweeps often contain ``jitter"—small deviations in speed and angle due to hand tremor or irregular motion. To evaluate our method's robustness, we introduced synthetic jitter to the simulated dataset. We define jitter as random noise added to the ground truth poses, scaled by a percentage of the total sweep extent (e.g., 2.5\% jitter corresponds to deviations up to 2.5\% of the total rotation range and image dimensions).

Since pose optimization is an optional component of our pipeline (\Cref{sec:pose_opt}), we study it as an ablation to assess how it improves robustness to motion jitter. Quantitative results for increasing jitter levels are summarized in \Cref{tab:results_jitter}.

As expected, introducing motion jitter degrades the performance of the fixed-pose baseline. At 2.5\% jitter—a level consistent with a reasonably steady hand—the test SSIM drops from 0.971 to 0.931. 

Activating learnable poses allows the model to correct for these deviations. Pose optimization significantly recovers performance, boosting Test SSIM from 0.931 back to 0.952, increasing PSNR and reducing LPIPS error. This demonstrates that our method can self-calibrate small motion artifacts without external tracking.

At extreme jitter levels (5\%), the optimization struggles to recover the correct geometry, likely getting trapped in local minima where the misalignment is too large for the gradients to guide the slices back to coherence. Interestingly, at this high noise level, the fixed-pose baseline outperforms the optimized version on the test set. This suggests that for highly erratic motion, the strong regularization of the smooth geometric prior is safer than attempting to optimize poses. However, for standard clinical sweeps, joint optimization provides a significant boost in fidelity.

\begin{table}[t]
\centering
\caption{Effect of learned poses on data affected by different percentages of jitter. }
\label{tab:results_jitter}
\setlength{\tabcolsep}{3pt}
\scriptsize
\begin{tabular}{l c c c c c c}
\hline
 & \multicolumn{3}{c}{\textbf{Training}} & \multicolumn{3}{c}{\textbf{Test}}\\
 &
SSIM & PSNR & LPIPS &
SSIM & PSNR & LPIPS \\
\hline
\multicolumn{7}{l}{\textbf{No Jitter}} \\
\hline
TVGS & \textbf{0.978} & 28.36 & \textbf{0.053} & \textbf{0.972} & 28.51 & \textbf{0.058}\\
TVGS w/o pose opt. & 0.975 & \textbf{28.54} & 0.057 & 0.971 & \textbf{28.83} & 0.059\\


\hline
\multicolumn{7}{l}{\textbf{2.5\% Jitter}} \\
\hline
TVGS & \textbf{0.950} & \textbf{23.06} & \textbf{0.099} & \textbf{0.952} & \textbf{23.88} & \textbf{0.089}\\
TVGS w/o pose opt. & 0.942 & 21.13 & 0.106 & 0.931 & 20.44 & 0.126\\


\hline
\multicolumn{7}{l}{\textbf{5\% Jitter}} \\
\hline
TVGS & \textbf{0.927} & \textbf{19.47} & \textbf{0.122} & 0.899 & 18.62 & 0.159\\
TVGS w/o pose opt. & 0.923 & 19.06 & 0.126 & \textbf{0.913} & \textbf{19.12} & \textbf{0.145}\\
\hline
\end{tabular}
\end{table}

\subsection{Reconstruction Quality}
\label{subsec:reconstruction_quality}

\subsubsection{View-Specific Reconstructions on Synthetic Data}
Based on insights gained from earlier experiments, we now evaluate the reconstruction performance of our method more closely.
We first evaluate our method on the synthetic dataset to validate geometric accuracy in a controlled environment. Using the default configuration of approximately 50,000 Gaussians, the optimization converges rapidly, requiring approximately 8 minutes for each view (sagittal and transversal).

\begin{figure}
\centering
\includegraphics[width=0.85\linewidth]{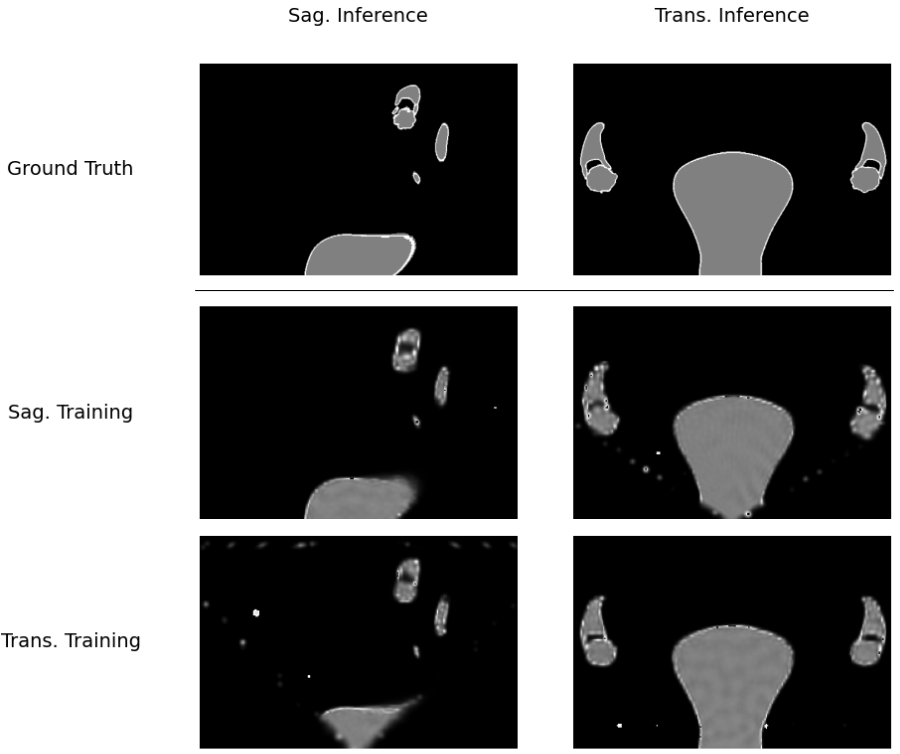}
\caption{Reconstruction results for sagittal and transversal sweeps. The top row shows results from sagittal training, the middle row from transversal training, and the bottom row shows the ground truth.}
\label{fig:synth_sweeps}
\end{figure}

\begin{figure}
\centering
\includegraphics[width=0.25\linewidth]{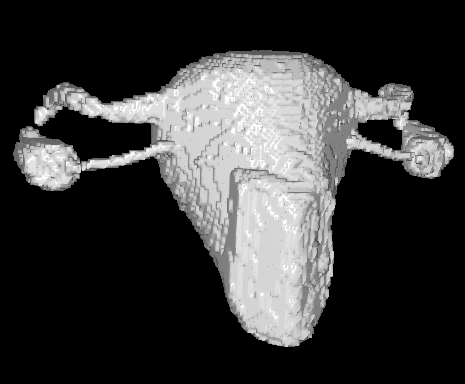}
\includegraphics[width=0.25\linewidth]{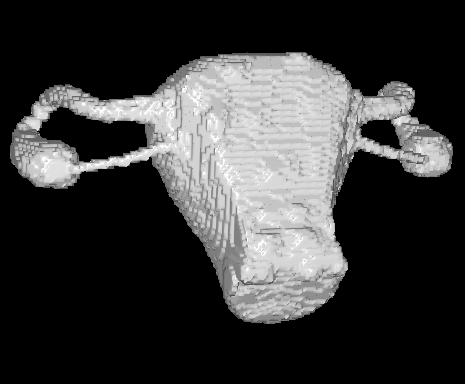}
\includegraphics[width=0.25\linewidth]{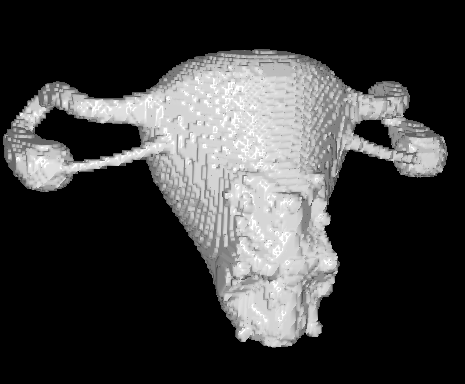}
\caption{3D reconstructions using different training settings. Left: sagittal only. Middle: transversal only. Right: both sagittal and transversal.}
\label{fig:3d_synth}
\end{figure}

\textbf{Qualitative Results.} \Cref{fig:synth_sweeps} illustrates the reconstruction fidelity. The model successfully captures the scene geometry, producing visually faithful renderings. We observe minor artifacts near the edges of structures, which we attribute to the sharp, non-smooth intensity transitions characteristic of binary segmentation masks used to generate the synthetic ground truth.
\Cref{fig:3d_synth} displays the resulting 3D volumetric representations. While individual sagittal or transversal sweeps capture the general structure, they exhibit cone-shaped artifacts due to the limited angular coverage of the ultrasound probe (spanning $\pm60$°). Combining both views minimizes these occlusions, resulting in a more complete 3D reconstruction.

\textbf{Comparison with Implicit Representations.}
Benchmarking unsupervised 3D US reconstruction is inherently challenging due to the lack of standard baselines that operate without external tracking. We compare our approach against ImplicitVol \cite{Yeung2024}, a state-of-the-art implicit neural representation method. To ensure a fair comparison in our tracking-free setting, we adapted ImplicitVol to use the slice poses estimated by our framework rather than pre-computed tracked poses.

Quantitative results are summarized in \Cref{tab:results_sim}. Our method achieves high structural similarity and peak signal-to-noise ratios, confirming the validity of the reconstruction. We observe that ImplicitVol achieves marginally higher quantitative scores (SSIM and PSNR) and lower perceptual error (LPIPS). This likely stems from the continuous nature of the MLPs in ImplicitVol, possessing an inductive bias towards smoothness well-suited for perfect, noise-free synthetic surfaces. In contrast, our discrete Gaussian representation may introduce minor high-frequency noise when approximating these perfectly smooth boundaries.

However, this slight difference in metric performance is outweighed by the computational efficiency of our approach. As shown in \Cref{tab:results_sim}, ImplicitVol requires approximately one hour to converge. In contrast, our Gaussian Splatting framework achieves comparable fidelity in just 8 to 16 minutes—representing an improvement in training speed of nearly an order of magnitude.

\begin{table}[t]
\centering
\caption{Quantitative comparison of our method with different number of Gaussians against ImplicitVol on simulated data.}
\label{tab:results_sim}
\setlength{\tabcolsep}{3pt}
\scriptsize
\begin{tabular}{l c c c c c c c}
\hline
 & \multicolumn{3}{c}{\textbf{Train}} & \multicolumn{3}{c}{\textbf{Test}} &  \\
 &
SSIM & PSNR & LPIPS &
SSIM & PSNR & LPIPS &
\textbf{Time} \\
\hline

\multicolumn{8}{l}{\textbf{Sagittal}} \\
\hline
ImplicitVol & \textbf{0.994} & \textbf{31.35} & \textbf{0.018} & \textbf{0.994} & \textbf{34.67} & \textbf{0.013} & 56m \\
TVGS (Ours) - 50k & 0.978 & 28.36 & 0.053 & 0.972 & 28.51 & 0.058 & \textbf{8m} \\
TVGS (Ours) - 100k & 0.981 & 29.00 & 0.047 & 0.974 & 29.02 & 0.051 & 16m \\

\hline
\multicolumn{8}{l}{\textbf{Transversal}} \\
\hline
ImplicitVol & \textbf{0.991} & \textbf{30.09} & \textbf{0.026} & \textbf{0.984} & \textbf{31.22} & \textbf{0.047} & 62m \\
TVGS (Ours) - 50k & 0.983 & 30.33 & 0.046 & 0.970 & 28.25 & 0.065 & \textbf{8m} \\
TVGS (Ours) - 100k & 0.986 & 31.47 & 0.039 & 0.972 & 28.55 & 0.054 & 16m \\
\hline
\end{tabular}
\end{table}



\subsubsection{Cross-View Generalization on Real Data}

Real US data introduces challenges compared to synthetic environments, particularly the lack of clear segmentation masks. Consequently, we evaluate reconstruction quality through both visual cross-view generalization and quantitative benchmarking against the adapted ImplicitVol baseline.

\begin{figure}
\centering
\includegraphics[width=0.85\linewidth]{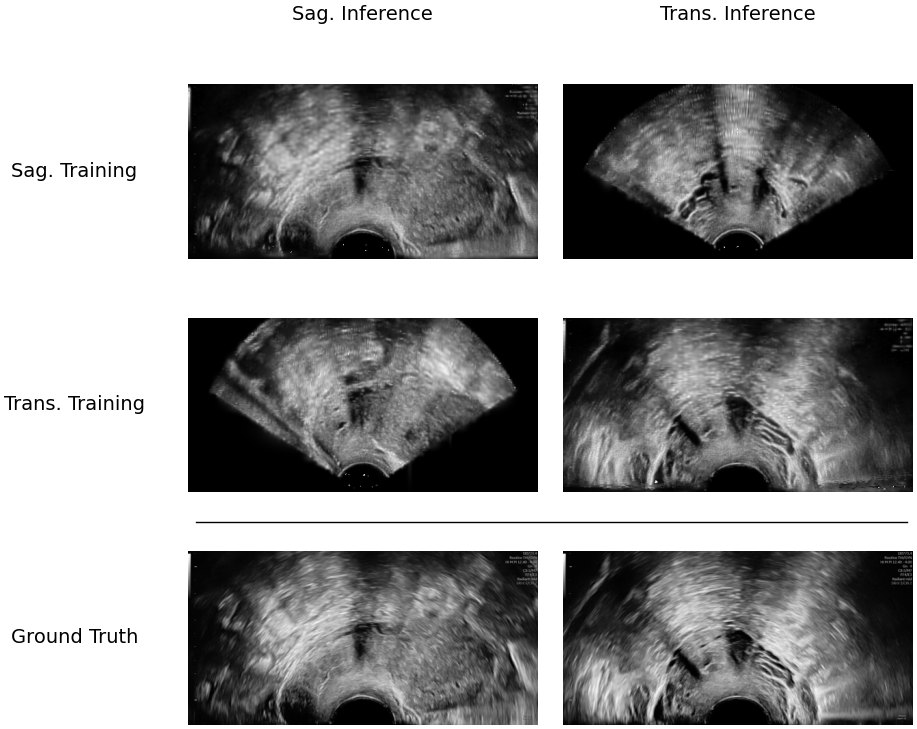}
\caption{Reconstructions from real data. Top: trained on sagittal view. Middle: trained on transversal view. Bottom: ground truth for comparison.}
\label{fig:gs_sweeps}
\end{figure}

\Cref{fig:gs_sweeps} demonstrates the qualitative capabilities of our model. Even when training is restricted to a single view (sagittal or transversal), the model generalizes well when tested on the orthogonal view. It captures anatomical features with high fidelity and handles difficult regions, such as areas obscured by fluid, with surprising accuracy.

To strictly quantify this performance, we compared our method against the ImplicitVol baseline on the real clinical data. As shown in \Cref{tab:results_real}, our approach's advantages are more pronounced here than in synthetic experiments. Our method outperforms the baseline across all metrics (SSIM, PSNR, LPIPS) for both sagittal and transversal sweeps. Notably, our approach achieves superior perceptual quality (indicated by lower LPIPS) and a drastic reduction in computational cost. While the implicit baseline requires over 3 hours to converge on real data, our method delivers superior results in just 11 to 16 minutes. This order-of-magnitude improvement in efficiency is critical for clinical adoption, where rapid feedback is essential.

\begin{table}[t]
\centering
\caption{Quantitative comparison of our method with different number of Gaussians against ImplicitVol on real patient data.}
\label{tab:results_real}
\scriptsize
\setlength{\tabcolsep}{2pt}
\begin{tabular}{lcccccccc}
\hline
 & \multicolumn{3}{c}{\textbf{Train}} & \multicolumn{3}{c}{\textbf{Test}} &  \\
 &
SSIM & PSNR & LPIPS &
SSIM & PSNR & LPIPS &
\textbf{Time} & \textbf{Memory} \\
\hline
\multicolumn{8}{l}{\textbf{Sagittal}} \\
\hline
ImplicitVol & 0.858 & 28.10 & 0.240 & 0.783 & 26.43 & 0.300 & 3.14h & 10GB\\

TVGS (Ours) - 50k & 0.901 & 31.38 & 0.188 & 0.845 & 28.51 & 0.223 & \textbf{11m} & \textbf{129MB}\\

TVGS (Ours) - 100k & \textbf{0.925} & \textbf{32.24} & \textbf{0.143} & \textbf{0.864} & \textbf{28.85} & \textbf{0.185} & 16m & 160MB\\
\hline

\multicolumn{8}{l}{\textbf{Transversal}} \\
\hline
ImplicitVol & 0.855 & 27.38 & 0.247 & 0.790 & 26.21 & 0.295 & 3.07h & 10GB\\

TVGS (Ours) - 50k & 0.897 & 30.81 & 0.194 & \textbf{0.847} & \textbf{28.44} & 0.223 & \textbf{11m} & \textbf{129MB}\\

TVGS (Ours) - 100k & \textbf{0.923} & \textbf{31.21} & \textbf{0.145} & 0.823 & 27.80 & \textbf{0.222} & 16m & 160MB\\
\hline

\end{tabular}
\end{table}

Despite the strong individual view reconstruction, combining multiple views remains a challenge. We observed alignment discrepancies between sagittal and transversal reconstructions, which complicates training on both views simultaneously. Addressing this misalignment is a key direction for future work; proper integration of both views could further refine slice position estimation, potentially moving beyond the simplified assumption of evenly spaced angles. In summary, our method demonstrates strong potential for reconstructing consistent and anatomically meaningful 3D representations from US sweeps, offering a scalable and fast alternative to existing implicit representations

\subsubsection{Effect of Slice Density}
\label{subsec:slice_density}
\begin{figure}[t!]
    \centering
    \includegraphics[width=0.75\linewidth]{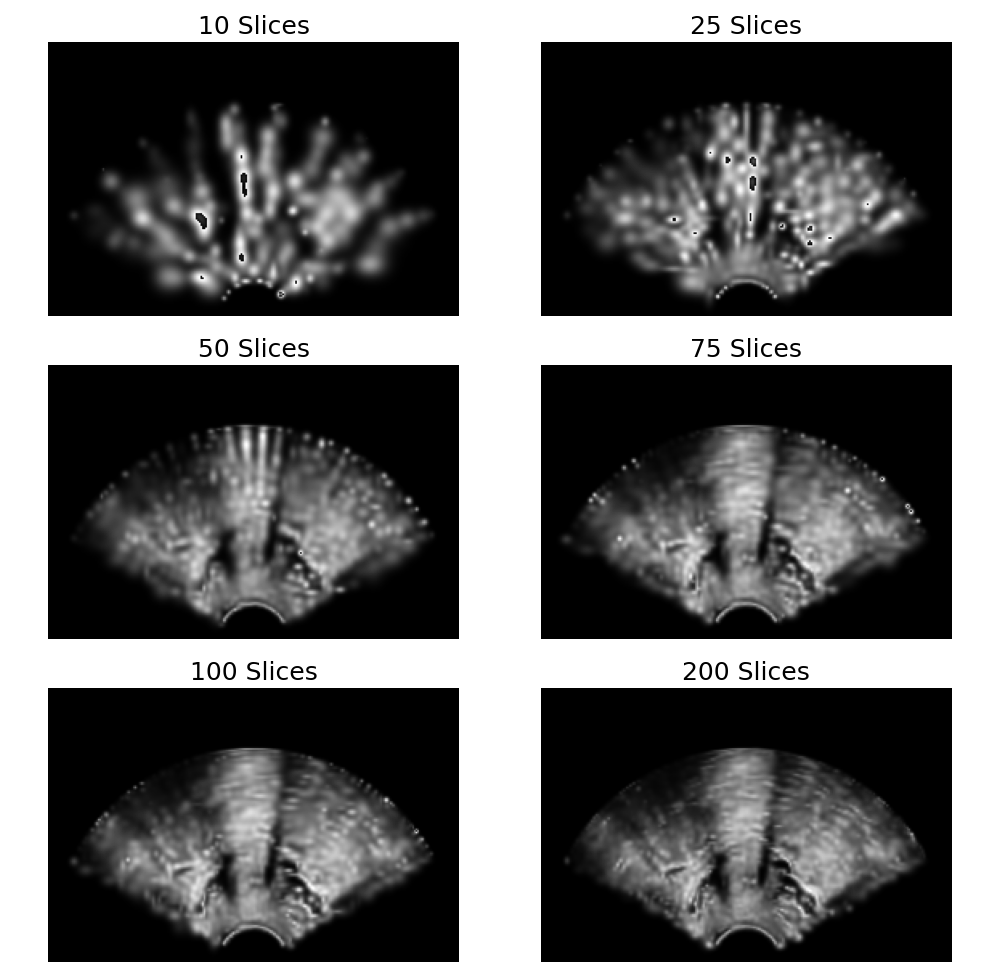}
    \caption{Reconstruction results for different amounts of slices.}
    \label{fig:num_slices_images}
\end{figure}

\begin{figure}[htbp]
    \centering
    \includegraphics[width=\linewidth]{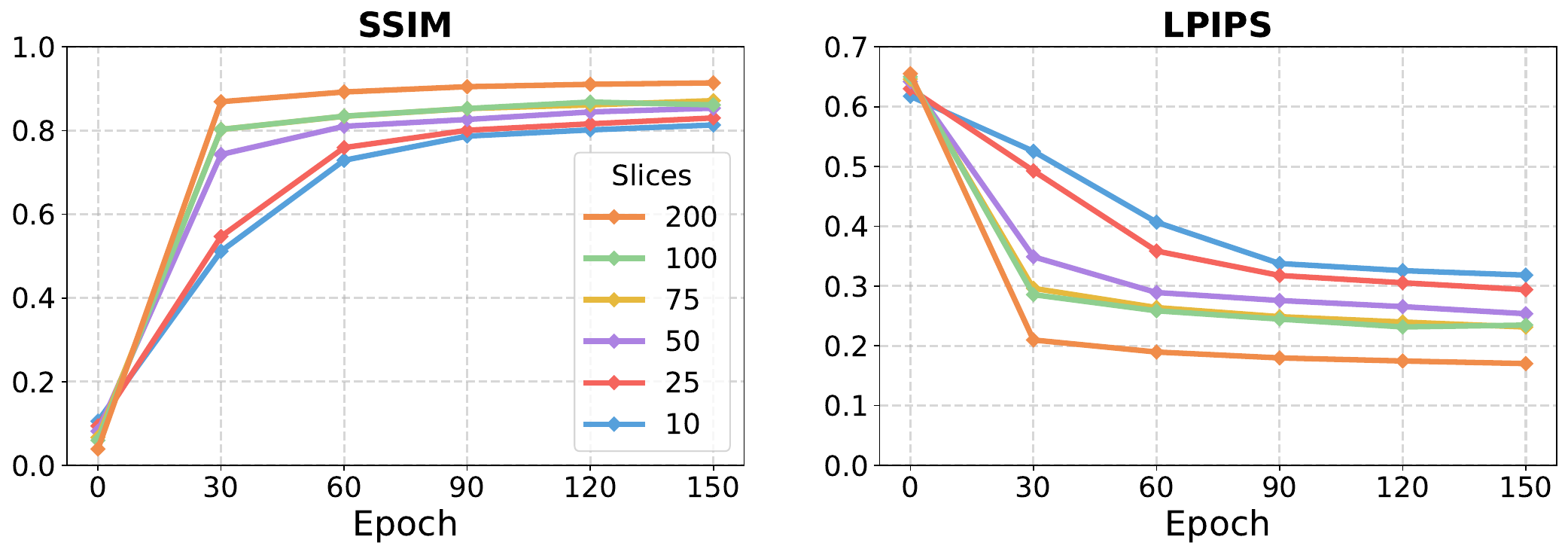}
    \caption{SSIM and LPIPS across epochs for different numbers of slices, same number of per-slice Gaussians.}
    \label{fig:slices_metrics}
\end{figure}

We also analyze the impact of slice density and Gaussian count. The number of slices acquired during a sweep significantly affects reconstruction quality. We systematically varied slice density—from 10 to 200 slices—to assess how it influences our model’s performance. As shown in \Cref{fig:num_slices_images}, reconstructions with very few slices (10–25) are blurry and lack anatomical detail, while quality improves substantially with 50 or more slices. Beyond 100 slices, gains diminish, suggesting that ultra-dense sampling offers limited additional benefit.
Quantitative trends in \Cref{fig:slices_metrics} confirm this: higher slice counts result in lower L1 loss and higher SSIM, and they also accelerate convergence. Conversely, sparse setups plateau at lower fidelity and slower training.
Our Gaussian primitives naturally compensate for moderate sparsity by expanding spatial support across neighboring slices. However, this ability is limited when data is extremely sparse, leading to smooth, under-detailed outputs. In contrast, denser setups allow Gaussians to reinforce structural coherence through natural overlap across slices.

\subsection{Rendering Speed}
\label{subsec:rendering_speed}
Finally, we evaluate the rendering performance by measuring the average frame rate under varying scene complexities. Specifically, we benchmark scenes containing different numbers of Gaussians to assess how the rendering speed scales with model size. The results, summarized in \Cref{tab:rendering_speed}, provide a quantitative comparison across these settings.

\begin{table}[ht]
    \centering
    \caption{Rendering speed different number of Gaussian primitives, for 10,000 iterations.}
    \label{tab:rendering_speed}
    \begin{tabular}{l l c c}
        \toprule
        \textbf{\# Gaussians} & \textbf{Sweep type} &  \textbf{Mean speed [ms]} & \textbf{FPS}\\
        \midrule
        \multirow{2}{*}{$\sim$ 10,000} & Transversal & 4.358 $\pm$ 0.040 & 229.47\\
                                  & Sagittal & 4.397 $\pm$ 0.031 & 227.40\\ 
        \midrule
        \multirow{2}{*}{$\sim$ 50,000} & Transversal & 18.142 $\pm$ 17.481 & 55.12\\
                                  & Sagittal & 18.182 $\pm$ 17.613 & 55.00\\ 
        \bottomrule
    \end{tabular}
\end{table}
\section{Discussion}

Our experiments highlight the sensitivity of the method to initialization. Gaussians placed near actual data slices converged faster and contributed more effectively to the final reconstruction, while randomly initialized primitives were often slow to optimize or entirely deactivated. Assigning higher learning rates to the mean parameters, compared to opacity or scale, helped reduce early deactivation and encouraged convergence.
Slice density was another important factor influencing performance. Denser sweeps led to higher-quality reconstructions, with fewer gaps and artifacts. Although adding more Gaussians can partially compensate for sparse data, this increases computational cost. This trade-off between reconstruction fidelity, data acquisition effort, and runtime is important to consider for clinical deployment.
The method also demonstrated promising generalization across orthogonal sweep directions. Volumes trained on sagittal sweeps could generalize well to transversal inputs and vice versa. However, combining multiple views simultaneously remains a challenge, particularly due to the need for accurate alignment. Our method does not currently perform registration, which can lead to visible inconsistencies when merging sweeps.

\noindent \textbf{Practical Limitations.} Our method relies on initialization from temporally ordered slices with uniform angular spacing. While our experiments on jitter (\Cref{tab:results_jitter}) demonstrate that joint pose optimization can effectively compensate for moderate motion irregularities and hand tremor, the method remains sensitive to rapid, large-scale changes in probe orientation that violate the initialization prior.
Furthermore, our rasterization framework utilizes a simplified additive model that neglects acoustic attenuation and shadowing. While this trade-off enables faster performance required for operator feedback, it limits the physical realism of the rendering compared to computationally intensive wave-propagation models.
Even with CUDA acceleration, the method remains computationally intensive. Near-real-time performance may be achievable with further optimization, but current runtimes are high, especially for large numbers of Gaussians or high-resolution data.
Finally, while we demonstrated feasibility using both synthetic and real datasets, broader clinical validation is needed. Differences in patient anatomy, imaging conditions, and probe operators were not extensively studied, and generalizing to these variables will be important before the method can be adopted in practice.

\noindent \textbf{Clinical Potential.} Despite these limitations, the method shows potential for improving clinical US workflows. During TVS examination the ability to generate accurate 3D reconstructions from freehand sweeps could support better visualization and localization of subtle anatomical abnormalities and enhance surgical planning. Additionally, real-time feedback during scanning could guide clinicians in covering relevant regions more thoroughly, potentially reducing missed findings making scanning less operator dependent.
The synthetic dataset and evaluation pipeline developed as part of this work may also serve as a foundation for future studies, offering reproducible benchmarks for US reconstruction research.

\section{Conclusion}
We have introduced a novel, unsupervised framework for reconstructing 3D anatomical volumes from freehand 2D transvaginal US sweeps, without requiring external tracking, ground truth trajectories, or pose supervision. 
By adapting Gaussian Splatting to the specific physics and geometry of US imaging, and designing a custom differentiable rasterizer, our method enables accurate, fast and memory-efficient volumetric reconstruction from sparse, uncalibrated slice data.
Through extensive experiments on both synthetic and real-world datasets, we demonstrated that the approach can achieve high-fidelity reconstructions, generalizes well across sweep orientations, and remains robust to various acquisition conditions. 
We showed that initialization, slice density, adaptive pose refinement, and the number of Gaussians are critical factors influencing performance, and proposed effective strategies such as density control for optimizing these parameters.
By eliminating the reliance on fixed geometric priors, the proposed learnable pose formulation proves that high-fidelity 3D US reconstruction can be achieved as a purely computational, self-correcting task.
While challenges remain—particularly regarding multi-view alignment and real-time performance—results suggest strong potential for clinical deployment.
Our method offers a scalable, hardware-free alternative to conventional 3D US systems and provides a foundation for future research in data-driven, volumetric reconstruction and AI-assisted gynecological imaging.

\printbibliography

\clearpage
\appendix

\subsection{Gradient Derivations}
\label{app:gradients}

This appendix provides the detailed derivations required for manually implementing the backward pass of the differentiable rasterizer. Since the forward pass is implemented using custom CUDA kernels, we cannot rely on autograd engines and must explicitly compute partial derivatives for each Gaussian parameter.

\subsubsection{Notation Recap}

For each pixel \( p \) at coordinate \( \mathbf{c}_p \in \mathbb{R}^3 \) and each Gaussian \( g \), we define:
\begin{align*}
    \mathbf{d}_{gp} &= \mathbf{c}_p - \mathbf{\mu}_g,\\
    e_{gp} &= -\frac{1}{2} \mathbf{d}_{gp}^\top \Sigma_g^{-1} \mathbf{d}_{gp},\\
    \alpha_{gp} &= o_g \exp(e_{gp}),\\
    I_p &= \sum_g \alpha_{gp} I_g
\end{align*}
The loss function \( L \) depends on rendered intensities \( I_p \), and we are given \( \frac{\partial L}{\partial I_p} \) from upstream layers.

\subsubsection{Gradient w.r.t.\ Gaussian Intensity \( I_g \)}

Since \( I_p = \sum_g \alpha_{gp} I_g \), differentiating w.r.t.\ \( I_g \) gives:
\begin{align*}
\frac{\partial I_p}{\partial I_g} &= \alpha_{gp} \\
\Rightarrow \frac{\partial L}{\partial I_g} &= \sum_p \frac{\partial L}{\partial I_p} \frac{\partial I_p}{\partial I_g} = \sum_p \frac{\partial L}{\partial I_p} \alpha_{gp}
\end{align*}

\subsubsection{Gradient w.r.t.\ Opacity \( o_g \)}

From \( \alpha_{gp} = o_g \exp(e_{gp}) \), we get:
\begin{align*}
\frac{\partial \alpha_{gp}}{\partial o_g} &= \exp(e_{gp}) \\
\Rightarrow \frac{\partial L}{\partial o_g} &= \sum_p \frac{\partial L}{\partial I_p} I_g \frac{\partial \alpha_{gp}}{\partial o_g} = \sum_p I_g \frac{\partial L}{\partial I_p} \exp(e_{gp})
\end{align*}

\subsubsection{Gradient w.r.t.\ Mean \( \mathbf{\mu}_g \)}

We use:
\[
\frac{\partial e_{gp}}{\partial \mathbf{d}_{gp}} = -\Sigma_g^{-1} \mathbf{d}_{gp},\quad
\frac{\partial \alpha_{gp}}{\partial \mathbf{d}_{gp}} = o_g \exp(e_{gp}) \left( -\Sigma_g^{-1} \mathbf{d}_{gp} \right)
\]

Then,
\[
\frac{\partial L}{\partial \mathbf{d}_{gp}} = \frac{\partial L}{\partial I_p} I_g \frac{\partial \alpha_{gp}}{\partial \mathbf{d}_{gp}} = -I_g \frac{\partial L}{\partial I_p} o_g \exp(e_{gp}) \Sigma_g^{-1} \mathbf{d}_{gp}
\]

Using \( \mathbf{d}_{gp} = \mathbf{c}_p - \mathbf{\mu}_g \), we get:
\[
\frac{\partial L}{\partial \mathbf{\mu}_g} = -\sum_p \frac{\partial L}{\partial \mathbf{d}_{gp}},\quad
\frac{\partial L}{\partial \mathbf{c}_p} = \sum_g \frac{\partial L}{\partial \mathbf{d}_{gp}}
\]

\subsubsection{Gradient w.r.t.\ Covariance \( \Sigma_g \)}

From:
\[
e_{gp} = -\frac{1}{2} \mathbf{d}_{gp}^\top \Sigma_g^{-1} \mathbf{d}_{gp},
\]
we use the identity:
\[
\frac{\partial e_{gp}}{\partial \Sigma_g} = \frac{1}{2} \Sigma_g^{-1} \mathbf{d}_{gp} \mathbf{d}_{gp}^\top \Sigma_g^{-1}
\]

Combining with the chain rule:
\begin{align*}
\frac{\partial L}{\partial e_{gp}} &= I_g \frac{\partial L}{\partial I_p} o_g \exp(e_{gp}) \\
\Rightarrow \frac{\partial L}{\partial \Sigma_g} &= \sum_p \frac{\partial L}{\partial e_{gp}} \frac{\partial e_{gp}}{\partial \Sigma_g} \\
&= \frac{1}{2} \sum_p I_g \frac{\partial L}{\partial I_p} o_g \exp(e_{gp}) \Sigma_g^{-1} \mathbf{d}_{gp} \mathbf{d}_{gp}^\top \Sigma_g^{-1}
\end{align*}

For covariance vector representation, we store only six unique elements:
\[
\mathrm{cov3D}_g = [\sigma_0,\sigma_1,\sigma_2,\sigma_3,\sigma_4,\sigma_5]
\Rightarrow
\Sigma_g =
\begin{bmatrix}
\sigma_0 & \sigma_1 & \sigma_2 \\
\sigma_1 & \sigma_3 & \sigma_4 \\
\sigma_2 & \sigma_4 & \sigma_5
\end{bmatrix}
\]

To compute \( \frac{\partial L}{\partial \sigma_i} \), we symmetrize the gradient:
\[
\frac{\partial L}{\partial \sigma_i} = 
\begin{cases}
(\tfrac{\partial L}{\partial \Sigma_g})_{ii} & \text{for diagonal } \sigma_i\\
\tfrac{1}{2} \left[ (\tfrac{\partial L}{\partial \Sigma_g})_{cd} + (\tfrac{\partial L}{\partial \Sigma_g})_{dc} \right] & \text{for off-diagonal } (c,d)
\end{cases}
\]

\subsubsection{Propagation to Scale and Rotation}
We use:
\begin{align*}
\Sigma_g &= M_g^\top M_g, \quad M_g = S_g R_g \\
\Rightarrow \frac{\partial L}{\partial M_g} &= 2 M_g \frac{\partial L}{\partial \Sigma_g}, \quad
\frac{\partial L}{\partial S_g} = \frac{\partial L}{\partial M_g} R_g^\top, \quad
\frac{\partial L}{\partial R_g} = S_g^\top\frac{\partial L}{\partial M_g}
\end{align*}

\subsubsection{Gradient w.r.t.\ Quaternion \( q_g \)}

We represent rotation using quaternions \( q_g = [q_r, q_i, q_j, q_k] \). The corresponding rotation matrix \( R(q_g) \) is defined analytically. To differentiate, we compute:
\[
\frac{\partial L}{\partial q_m} = \sum_{u,v} \frac{\partial L}{\partial [R_g]_{uv}} \frac{\partial [R(q_g)]_{uv}}{\partial q_m}, \quad m \in \{r, i, j, k\}
\]

To backpropagate through quaternion normalization \( q_g = \tilde{q}_g / \|\tilde{q}_g\| \), we use:
\[
\frac{\partial L}{\partial \tilde{q}_g} = \frac{1}{\|\tilde{q}_g\|} \left( I - q_g q_g^\top \right) \frac{\partial L}{\partial q_g}
\]

\subsection{Initialization Schemes}
\label{app:init}
We define the two initialization strategies used for setting Gaussian means \( \mathbf{\mu}_g \):

\subsubsection{On-slice initialization.}
Assuming a known slice pose with rotation \( R_S \in \mathbb{R}^{3\times3} \) and translation \( \mathbf{t}_S \in \mathbb{R}^3 \), we define a 2D reference grid:
\[
\mathcal{P} = \{ (x_i, y_j, 0) \in \mathbb{R}^3 \mid x_i \in [a_x, b_x],\; y_j \in [a_y, b_y] \}
\]
with uniform grid spacing. The 3D Gaussian centers are then computed via:
\[
\mathbf{\mu}_{ij} = R_S (p_{ij} + \mathbf{t}_S), \quad p_{ij} \in \mathcal{P}
\]

\subsubsection{Uniform sampling.}
If no pose is known, we sample:
\[
\mathbf{\mu}_g = \mathbf{a} + (\mathbf{b} - \mathbf{a}) \odot \mathbf{\xi}_g, \quad \mathbf{\xi}_g \sim \mathcal{U}[0,1]^3
\]
where \( \mathbf{a}, \mathbf{b} \in \mathbb{R}^3 \) define the bounding box.

\begin{figure}
    \centering
    \includegraphics[width=.49\linewidth]{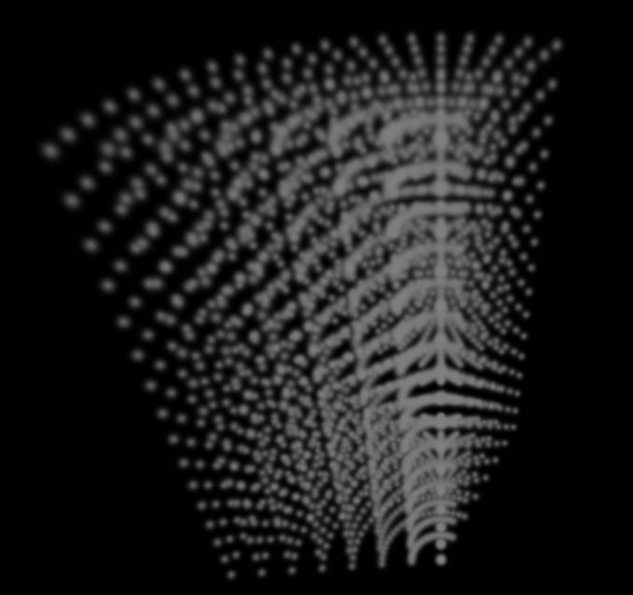}
    \includegraphics[width=.49\linewidth]{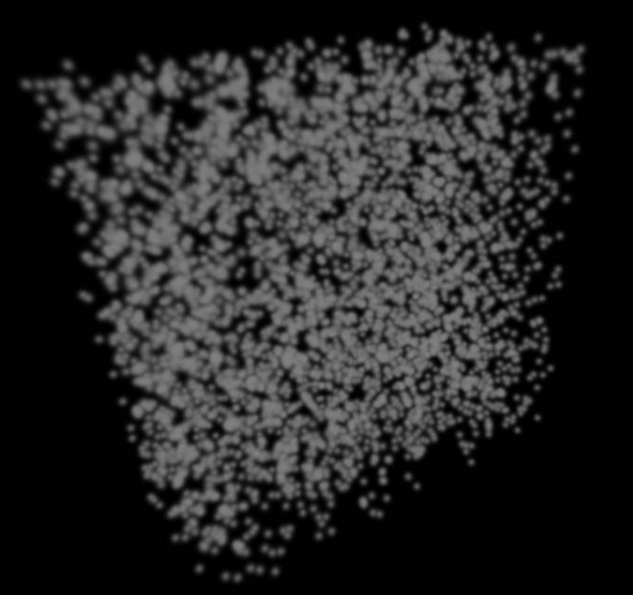}
    \caption{Left: On-Slice initialization for a sagittal sweep. Right: Random initialization for a sagittal sweep.}
    \label{fig:init_strats}
\end{figure}

A visual comparison of these two initialization strategies is shown in Fig.~\ref{fig:init_strats}. 
On-slice initialization produces structured, well-aligned Gaussians from the start, whereas uniform sampling yields a diffuse cloud of Gaussians that must be refined during training. 

\subsection{Inactive Gaussians and Density Control}
\label{app:density}

Our density control mechanism involves periodic pruning and reinitialization of Gaussians based on their contribution metric \( m_g = |I_g \cdot o_g| \). 
Below are the key implementation details:

\subsubsection{Pruning Schedule.}  
Every 10 epochs, we evaluate all Gaussians and mark those with:
\[
m_g = |I_g \cdot o_g| < \epsilon, \quad \text{with } \epsilon = 10^{-4}
\]
as inactive. These are removed from the parameter set.

\subsubsection{Reinitialization.}  
For each removed Gaussian, a new one is inserted. The new mean \( \mathbf{\mu}_g \) is sampled uniformly within the current bounding box of active Gaussian means:
\[
\mathbf{\mu}_g \sim \mathcal{U}[\mathbf{\mu}_{\min}, \mathbf{\mu}_{\max}]
\]
All other parameters are initialized using the defaults described in Section~\ref{sec:initialization}.

\subsubsection{Optional Filters.}  
To prevent rapid re-deactivation, newly inserted Gaussians can be given a brief "grace period" (e.g., skipped in the next one or two pruning steps).
This lightweight mechanism stabilizes training and improves efficiency, particularly in early epochs where many Gaussians otherwise become stuck.

Tables \ref{tab:lr_means} and \ref{tab:lr} show the search results for the alternative method proposed for inactive Gaussians. 

\begin{table}[h] 
    \centering
    \caption{L1 and SSIM metrics on different learning rates for means}
    \label{tab:lr_means}
    \begin{tabular*}{0.49\textwidth}{@{\extracolsep{\fill}}lcc|cc|cc}
        \toprule
         & \multicolumn{2}{c}{0.15} & \multicolumn{2}{c}{0.2} & \multicolumn{2}{c}{0.25} \\
         \cmidrule(lr){2-3} \cmidrule(lr){4-5} \cmidrule(lr){6-7}
        \textbf{Metric} & \textbf{L1} & \textbf{SSIM} & \textbf{L1} & \textbf{SSIM} & \textbf{L1} & \textbf{SSIM}\\
        \midrule
        Means & 0.0201 & 0.8642 & \textbf{0.0189} & 0.8692 & 0.201 & \textbf{0.8700}\\
        \bottomrule
    \end{tabular*} 
\end{table}

\begin{table}[h] 
    \centering
    \caption{L1 and SSIM metrics on different learning rates}
    \label{tab:lr}
    \begin{tabular*}{0.49\textwidth}{@{\extracolsep{\fill}}lcc|cc|cc}
        \toprule
         & \multicolumn{2}{c}{0.005} & \multicolumn{2}{c}{0.01} & \multicolumn{2}{c}{0.05} \\
         \cmidrule(lr){2-3} \cmidrule(lr){4-5} \cmidrule(lr){6-7}
        \textbf{Metric} & \textbf{L1} & \textbf{SSIM} & \textbf{L1} & \textbf{SSIM} & \textbf{L1} & \textbf{SSIM} \\
        \midrule
        Opacities      & 0.0228 & 0.8451 & \textbf{0.0213} & 0.8485 & 0.0221 & \textbf{0.8518}\\
        Scales         & 0.0171 & 0.8671 & \textbf{0.0169} & \textbf{0.8680} & 0.0221 & 0.8518\\
        Intensities    & \textbf{0.0194} & 0.8540 & 0.0195 & \textbf{0.8546} & 0.0221 & 0.8518\\
        \bottomrule
    \end{tabular*} 
\end{table}

\subsection{Spatial convergence}

\begin{figure}
    \centering
    \includegraphics[width=.9\linewidth]{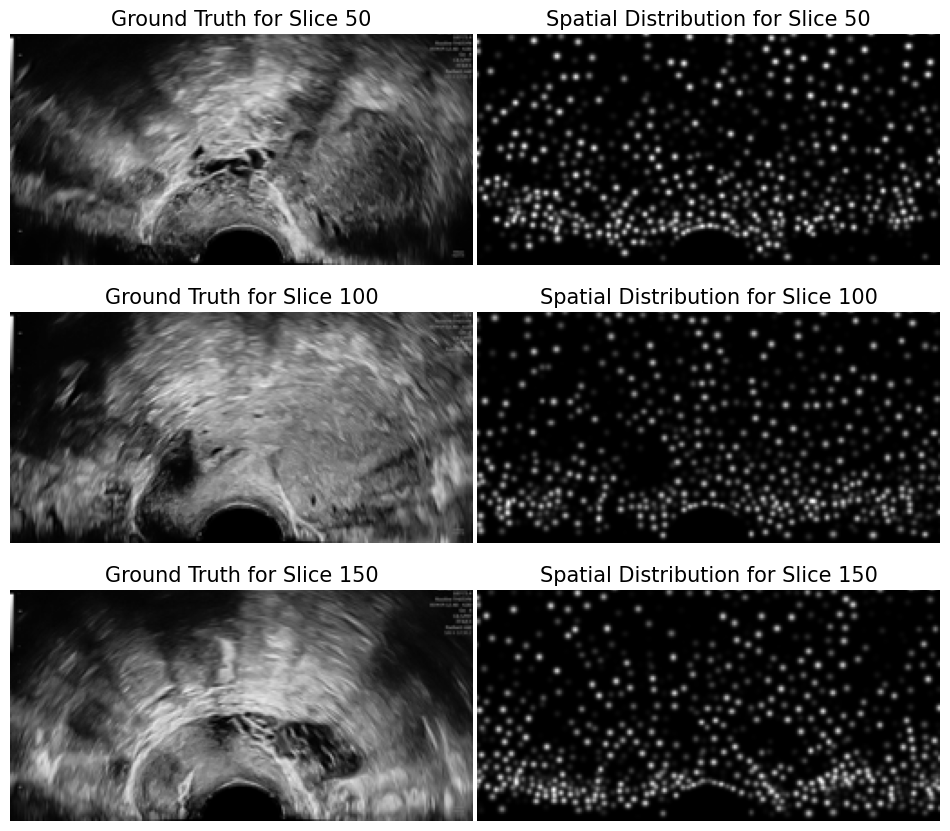}
    \caption{Distribution of Gaussian primitives for a scene.}
    \label{fig:gspace}
\end{figure}

Fig. \ref{fig:gspace} visualizes the spatial distribution of Gaussians after optimization. For clarity, all opacities and intensities are set to 1.0, and Gaussian scales are reduced by a factor of 0.2. The visualizations show that Gaussians concentrate in high-intensity areas corresponding to meaningful anatomical structures, while low-intensity regions remain largely empty—reflecting the model’s ability to focus its representational capacity where it matters most.

\begin{figure}
    \centering
    \includegraphics[width=.85\linewidth]{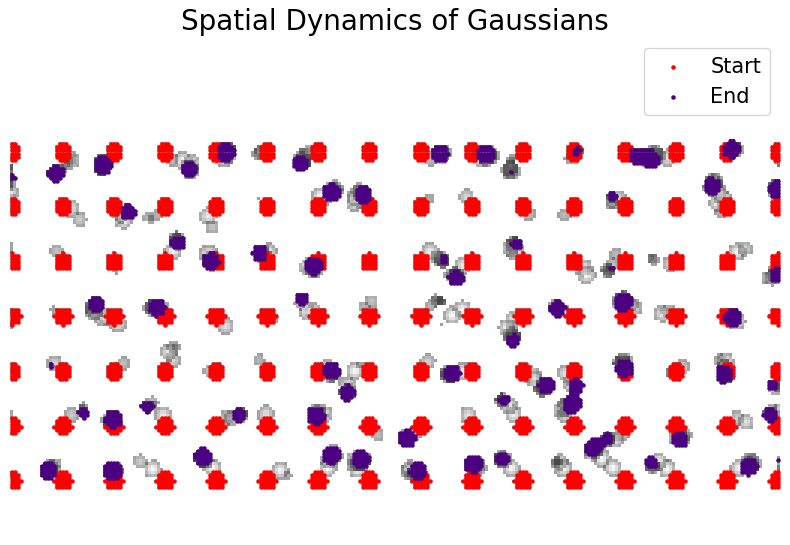}
    \includegraphics[width=.85\linewidth]{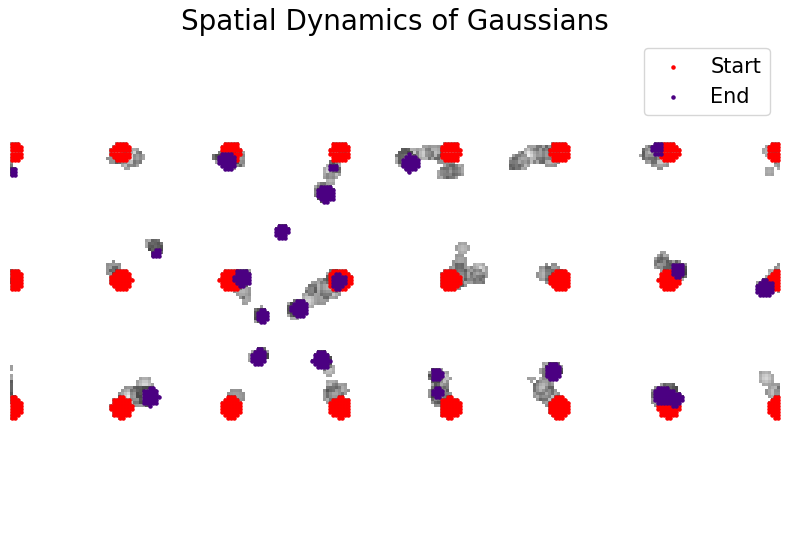}
    \caption{Trajectory of Gaussian centers during training for the same slice, comparing different quantities of Gaussians. Grey traces indicate movement paths. Some Gaussians appear only at the start or end of the trajectory within the slice, suggesting they originated from or moved to adjacent slices.}
    \label{fig:gs_dyn}
\end{figure}

Fig. \ref{fig:gs_dyn} illustrates how Gaussian centers evolve during training, comparing a sparse configuration ($\approx 24$ Gaussians per slice) to a denser one ($\approx112$ per slice). With fewer Gaussians, individual movements are larger and more distinct, as each primitive covers more space. In denser settings, movement is subtler due to increased redundancy.

Notably, Gaussians exhibit dynamic reallocation: some exit the visualized slice while others enter or converge toward high-intensity regions. This behavior suggests an adaptive mechanism that reallocates primitives toward regions of higher relevance or deactivates them when unnecessary, enabling efficient use of limited representational resources.

\subsection{Implementation Optimizations}
\label{app:performance}

Efficient training of the differentiable rasterizer requires high-performance implementation of both the forward and backward passes due to the dense, per-pixel evaluation of contributions from thousands of 3D Gaussians. 
This section describes the key CUDA-level optimizations applied to ensure computational feasibility and scalability.

\subsubsection{Precomputation of Reusable Quantities}
The backward pass relies heavily on values such as \( \Sigma_g^{-1} \), which are reused across all pixels influenced by a given Gaussian. 
To avoid redundant computation, we launch a dedicated kernel that precomputes \( \Sigma_g^{-1} \) for all Gaussians and stores the upper-triangular components in a compact 6-element representation. 
These cached inverses are then accessed during both forward and backward passes to evaluate \( e_{gp} \), \( \alpha_{gp} \), and their gradients.
Similarly, we precompute \( \Sigma_g^{-1} \mathbf{d}_{gp} \) wherever possible to avoid recomputation during the accumulation of gradient terms.
This significantly reduces arithmetic overhead in the main rasterization kernels.

\subsubsection{Shared Memory Tiling}
A major bottleneck in GPU rasterization is the repeated access to Gaussian parameters from global memory. 
To mitigate this, we divide the set of Gaussians into tiles of size \( T \) (e.g., 64).
Each CUDA thread block loads one tile of Gaussian data (means, scales, quaternions, opacities, intensities) into shared memory.
Threads within the block synchronize and reuse the shared memory tile for computing contributions and gradients for their assigned pixels.
This results in reduction in global memory bandwidth usage, better memory coalescing and cache locality and increased parallel efficiency and occupancy.

\subsubsection{Symmetric Covariance Storage and Update}
Each 3D Gaussian covariance matrix \( \Sigma_g \) is symmetric. 
During the backward pass, we compute \( \frac{\partial L}{\partial \Sigma_g} \) as a full matrix.
Each symmetric entry (e.g., \( \sigma_1 \), appearing in both \( (0,1) \) and \( (1,0) \)) receives contributions from both locations and is updated using their average.
This saves memory, reduces write operations, and simplifies gradient accumulation logic.

\subsubsection{Matrix-Free Operations}

To minimize expensive matrix multiplications and inversions, we unroll expressions like \( \mathbf{d}_{gp}^\top \Sigma_g^{-1} \mathbf{d}_{gp} \) into explicit component-wise dot products using precomputed inverse covariances.
Gradients involving matrix–vector products are similarly computed via direct arithmetic to avoid launching linear algebra kernels.
This low-level optimization is especially effective for backward steps where every saved FLOP improves scalability.

\subsubsection{Compiler-Level Optimizations}
To further improve performance, we apply the following compiler flags during kernel compilation
\begin{itemize}
    \item \texttt{-O3}: Enables aggressive optimizations like loop unrolling, constant folding, and function inlining.
    \item \texttt{--use\_fast\_math}: Activates fast approximations for transcendental functions (e.g., \texttt{exp}, \texttt{sqrt}), which are sufficient for training purposes and significantly faster.
\end{itemize}
In practice, these flags reduce kernel runtime by 20–40\% without compromising training stability.

\subsubsection{Quantitative Impact}
\begin{figure}
    \centering
    \includegraphics[width=0.9\linewidth]{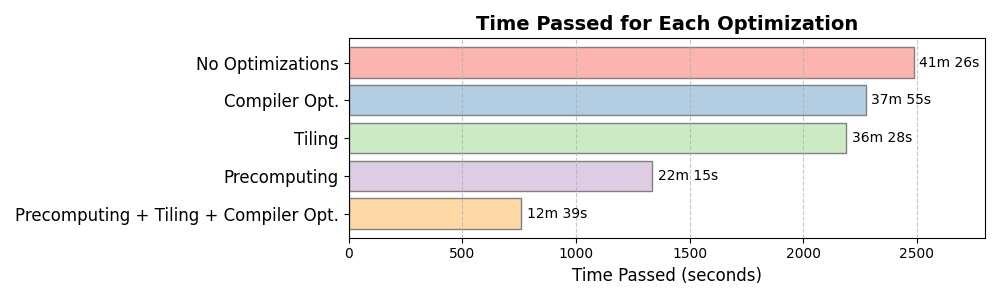}
    \caption{Effect of cumulative optimization strategies on training time.}
    \label{fig:optimizations}
\end{figure}

Fig.~\ref{fig:optimizations} shows the training time required to fit a fixed US volume as each optimization is introduced cumulatively. 
The baseline PyTorch implementation (no tiling, no precomputation, no matrix unrolling) takes approximately 18 hours. 
Our optimized CUDA implementation reduces this to under 2 hours for a scene with over 50,000 Gaussians.

\subsection{Extended Results}

This section presents a detailed quantitative breakdown of the reconstruction performance across individual subjects. While the main text reports aggregated statistics, the tables below illustrate the consistency of our method across different patient scans. We compare the baseline ImplicitVol approach against our differentiable rasterization method configured with 50,000 and 100,000 primitives. As shown in the per-patient breakdowns, our approach achieves competitive similarity metrics (SSIM, PSNR) and perceptual quality (LPIPS) while consistently reducing the computational burden from approximately 3 hours to under 20 minutes per volume.

\begin{table}[t]
\centering
\caption{Quantitative results in the case of simulated data.}
\label{tab:results}
\scriptsize
\begin{tabular}{l c c c c c c c}
\hline
 & \multicolumn{3}{c}{\textbf{Training}} & \multicolumn{3}{c}{\textbf{Test}} &  \\
\textbf{Pat. ID} &
SSIM & PSNR & LPIPS &
SSIM & PSNR & LPIPS &
\textbf{Time} \\
\hline

\multicolumn{8}{l}{\textbf{Sag.}} \\
\hline
ImplicitVol & 0.9944 & 31.35 & 0.0182 & 0.9936 & 34.67 & 0.0132 & 56m \\
Ours - 50k & 0.9760 & 28.61 & 0.0556 & 0.9301 & 28.35 & 0.1284 & 8m \\
Our - 100k & 0.9792 & 29.41 & 0.0502 & 0.9218 & 28.53 & 0.1350 & 16m \\

\hline
\multicolumn{8}{l}{\textbf{Trans.}} \\
\hline
ImplicitVol & 0.9910 & 30.09 & 0.0255 & 0.9835 & 31.22 & 0.0468 & 62m \\
Ours - 50k & 0.9834 & 30.36 & 0.0461 & 0.8770 & 27.98 & 0.2185 & 8m \\
Ours - 100k & 0.9865 & 31.75 & 0.0394 & 0.8378 & 27.77 & 0.2319 & 16m \\

\hline
\end{tabular}
\end{table}

\begin{table}[t]
\centering
\tiny
\caption{Quantitative results for different patients in the case of ImplicitVol on real data.}
\label{tab:results}
\begin{tabular}{l c c c c c c c}
\hline
 & \multicolumn{3}{c}{\textbf{Training}} & \multicolumn{3}{c}{\textbf{Test}} &  \\
\textbf{Pat. ID} &
SSIM & PSNR & LPIPS &
SSIM & PSNR & LPIPS &
\textbf{Time} \\
\hline

\multicolumn{8}{l}{\textbf{Sag.}} \\
\hline
152 & 0.8459 & 27.11 & 0.2599 & 0.7730 & 25.62 & 0.3145 & 3.15h \\
154 & 0.8364 & 27.44 & 0.2634 & 0.7709 & 26.28 & 0.3081 & 3.34h \\
159 & 0.8480 & 27.06 & 0.2506 & 0.7731 & 25.61 & 0.3122 & 3.12h \\
231 & 0.8397 & 27.53 & 0.2533 & 0.7570 & 25.95 & 0.3135 & 3.03h \\
232 & 0.8970 & 28.90 & 0.2151 & 0.8229 & 27.18 & 0.2831 & 3.13h \\
238 & 0.8922 & 29.87 & 0.2029 & 0.8204 & 27.79 & 0.2720 & 3.02h \\
274 & 0.8600 & 28.66 & 0.2262 & 0.7806 & 26.53 & 0.2963 & 3.17h \\
306 & 0.8317 & 27.45 & 0.2566 & 0.7573 & 25.92 & 0.3028 & 3.13h \\
324 & 0.8924 & 29.15 & 0.2127 & 0.8126 & 26.87 & 0.2835 & 3.13h \\
326 & 0.8357 & 27.80 & 0.2611 & 0.7670 & 26.54 & 0.3156 & 3.16h \\

\hline
\multicolumn{8}{l}{\textbf{Trans.}} \\
\hline
152 & 0.8537 & 26.81 & 0.2591 & 0.7868 & 25.73 & 0.3076 & 3.07h \\
154 & 0.8154 & 26.07 & 0.2810 & 0.7588 & 25.33 & 0.3144 & 3.13h \\
159 & 0.8595 & 27.57 & 0.2491 & 0.8080 & 26.33 & 0.2908 & 3.08h \\
231 & 0.8307 & 26.19 & 0.2660 & 0.7582 & 25.23 & 0.3157 & 3.09h \\
232 & 0.9144 & 28.90 & 0.2059 & 0.8521 & 27.12 & 0.2778 & 3.05h \\
238 & 0.8762 & 28.68 & 0.2218 & 0.8235 & 27.55 & 0.2629 & 3.06h \\
274 & 0.8556 & 27.96 & 0.2340 & 0.7898 & 26.87 & 0.2793 & 3.07h \\
306 & 0.8237 & 26.21 & 0.2644 & 0.7491 & 25.40 & 0.3056 & 3.09h \\
324 & 0.8879 & 29.18 & 0.2226 & 0.8128 & 27.25 & 0.2811 & 2.97h \\
326 & 0.8285 & 26.18 & 0.2652 & 0.7647 & 25.27 & 0.3120 & 3.11h \\


\hline
\end{tabular}
\end{table}

\begin{table}[t]
\centering
\tiny
\caption{Quantitative results for different patients in the case of Ours-50k on real data. Trained on 150 epochs}
\label{tab:results}
\begin{tabular}{l c c c c c c c}
\hline
 & \multicolumn{3}{c}{\textbf{Training}} & \multicolumn{3}{c}{\textbf{Test}} &  \\
\textbf{Pat. ID} &
SSIM & PSNR & LPIPS &
SSIM & PSNR & LPIPS &
\textbf{Time} \\
\hline

\multicolumn{8}{l}{\textbf{Sag.}} \\
\hline
152 & 0.8903 & 30.40 & 0.2099 & 0.8303 & 27.81 & 0.2543 & 11m \\
154 & 0.8866 & 30.77 & 0.2088 & 0.8412 & 28.74 & 0.2299 & 11m \\
159 & 0.8899 & 30.61 & 0.1970 & 0.8378 & 28.28 & 0.2309 & 11m \\
231 & 0.8901 & 30.97 & 0.1991 & 0.8279 & 28.39 & 0.2428 & 11m \\
232 & 0.9248 & 32.24 & 0.1723 & 0.8657 & 28.82 & 0.2106 & 11m \\
238 & 0.9200 & 33.03 & 0.1603 & 0.8658 & 29.08 & 0.1964 & 11m \\
274 & 0.9040 & 31.83 & 0.1735 & 0.8484 & 28.61 & 0.2103 & 11m \\
306 & 0.8875 & 30.59 & 0.1910 & 0.8298 & 27.99 & 0.2207 & 11m \\
324 & 0.9260 & 32.68 & 0.1698 & 0.8616 & 28.54 & 0.2107 & 11m \\
326 & 0.8913 & 30.72 & 0.1934 & 0.8442 & 28.79 & 0.2209 & 11m \\

\hline
\multicolumn{8}{l}{\textbf{Trans.}} \\
\hline
152 & 0.8967 & 30.16 & 0.2032 & 0.8431 & 27.94 & 0.2321 & 11m \\
154 & 0.8696 & 29.41 & 0.2196 & 0.8250 & 27.69 & 0.2453 & 11m \\
159 & 0.8998 & 30.88 & 0.1968 & 0.8609 & 28.83 & 0.2224 & 11m \\
231 & 0.8834 & 29.67 & 0.2039 & 0.8237 & 27.67 & 0.2352 & 11m \\
232 & 0.9329 & 32.44 & 0.1708 & 0.8854 & 29.31 & 0.2112 & 11m \\
238 & 0.9129 & 32.72 & 0.1748 & 0.8707 & 30.14 & 0.1931 & 11m \\
274 & 0.8981 & 31.01 & 0.1859 & 0.8470 & 28.67 & 0.2138 & 11m \\
306 & 0.8778 & 29.90 & 0.1978 & 0.8121 & 27.26 & 0.2307 & 11m \\
324 & 0.9227 & 32.76 & 0.1818 & 0.8628 & 28.80 & 0.2122 & 11m \\
326 & 0.8800 & 29.11 & 0.2041 & 0.8425 & 28.12 & 0.2357 & 11m \\


\hline
\end{tabular}
\end{table}

\begin{table}[t]
\centering
\tiny
\caption{Quantitative results for different patients in the case of Ours-100k on real data. Trained on 150 epochs}
\label{tab:results}
\begin{tabular}{l c c c c c c c}
\hline
 & \multicolumn{3}{c}{\textbf{Training}} & \multicolumn{3}{c}{\textbf{Test}} &  \\
\textbf{Pat. ID} &
SSIM & PSNR & LPIPS &
SSIM & PSNR & LPIPS &
\textbf{Time} \\
\hline

\multicolumn{8}{l}{\textbf{Sag.}} \\
\hline
152 & 0.9157 & 31.21 & 0.1638 & 0.8471 & 28.12 & 0.2118 & 16m \\
154 & 0.9157 & 31.80 & 0.1590 & 0.8613 & 29.19 & 0.1879 & 16m \\
159 & 0.9161 & 31.48 & 0.1523 & 0.8588 & 28.68 & 0.1877 & 16m \\
231 & 0.9167 & 31.77 & 0.1526 & 0.8471 & 28.86 & 0.2055 & 16m \\
232 & 0.9415 & 32.88 & 0.1326 & 0.8792 & 29.00 & 0.1761 & 16m \\
238 & 0.9368 & 33.69 & 0.1253 & 0.8781 & 29.30 & 0.1656 & 16m \\
274 & 0.9281 & 32.60 & 0.1306 & 0.8644 & 28.86 & 0.1772 & 16m \\
306 & 0.9182 & 31.65 & 0.1406 & 0.8542 & 28.40 & 0.1791 & 16m \\
324 & 0.9439 & 33.63 & 0.1277 & 0.8771 & 28.82 & 0.1746 & 16m \\
326 & 0.9199 & 31.71 & 0.1446 & 0.8683 & 29.28 & 0.1800 & 16m \\

\hline
\multicolumn{8}{l}{\textbf{Trans.}} \\
\hline
152 & 0.9215 & 30.97 & 0.1574 & 0.8602 & 28.26 & 0.1926 & 16m \\
154 & 0.9027 & 30.38 & 0.1679 & 0.8409 & 27.89 & 0.2080 & 16m \\
159 & 0.9227 & 31.69 & 0.1523 & 0.8802 & 29.34 & 0.1833 & 16m \\
231 & 0.9083 & 30.26 & 0.1623 & 0.8323 & 27.63 & 0.2066 & 16m \\
232 & 0.9448 & 32.80 & 0.1369 & 0.8908 & 29.40 & 0.1804 & 16m \\
238 & 0.9348 & 33.71 & 0.1333 & 0.8881 & 30.53 & 0.1578 & 16m \\
274 & 0.9224 & 31.87 & 0.1430 & 0.8620 & 28.93 & 0.1834 & 16m \\
306 & 0.9097 & 30.94 & 0.1506 & 0.8324 & 27.63 & 0.1886 & 16m \\
324 & 0.9399 & 33.48 & 0.1413 & 0.8719 & 28.93 & 0.1823 & 16m \\
326 & 0.9077 & 29.77 & 0.1610 & 0.8632 & 28.39 & 0.1921 & 16m \\


\hline
\end{tabular}
\end{table}


\end{document}